\documentclass[11pt]{article}
\usepackage{fullpage}
\usepackage{graphicx}
\usepackage{xspace}
\usepackage{url}
\usepackage{amssymb}
\usepackage{latexsym}
\usepackage{amsthm}
\usepackage{amsmath}
\usepackage{amstext}
\usepackage{amstext}
\usepackage{algorithmic}
\usepackage{enumitem}
\usepackage{algorithm}
\usepackage{eucal}
\usepackage{tabularx}
\usepackage{vmargin}
 \setmarginsrb{1in}{1in}{1in}{1in}{0pt}{0pt}{0pt}{7mm}
\usepackage{comment}
\usepackage{caption}
\usepackage{color}
\newcommand{\xs}{\ \ \ }

\newcommand{\executeiffilenewer}[3]{%
\ifnum\pdfstrcmp{\pdffilemoddate{#1}}%
{\pdffilemoddate{#2}}>0%
{\immediate\write18{#3}}\fi%
} 

\newcommand{\svg}[2]{\def\svgwidth{#1}%
\executeiffilenewer{#2.svg}{#2.pdf}%
{inkscape -z -D --file=#2.svg %
--export-pdf=#2.pdf --export-latex}%
{\input{#2.pdf_tex}}}

\newcommand{\M}{\mathcal{M}}
\newcommand{\D}{\mathcal{D}}
\newcommand{\T}{\mathcal{T}}
\newcommand{\R}{\mathcal{R}}
\newcommand{\I}{\mathcal{I}}
\newcommand{\np}{\mathsf{NP}}
\newcommand{\wone}{\mathsf{W[1]}}
\newcommand{\nat}{\mathbb{N}}
\newcommand{\pr}{\textup{pr}}
\newcommand{\view}{\textup{view}}
\newcommand{\Int}{\textup{Int}}

\newcommand{\vname}[1]{\textsf{#1}}
\newcommand{\prob}[3]{
\begin{center}\fbox{\parbox{0.9\linewidth}{
#1\\
\begin{tabularx}{\linewidth}{rX}
\textbf{Input:} & #2\\
\textbf{Question:} & #3
\end{tabularx}
}}
\end{center}
}

\newif\ifabstract
\abstracttrue

\abstractfalse 

\newif\iffull
\ifabstract 
\usepackage{mathptmx}
\usepackage[nomarkers,nolists]{endfloat}

\DeclareDelayedFloat{algorithm}{Algorithms}
 \setmarginsrb{0.97in}{0.97in}{0.97in}{0.97in}{0pt}{0pt}{0pt}{7mm}
\renewcommand{\baselinestretch}{0.97}
\fullfalse \else \fulltrue  \fi

\floatname{algorithm}{Algorithm}
\renewcommand{\subset}{\subseteq}

\newtheorem{theorem}{Theorem}[section]
\newtheorem{lemma}[theorem]{Lemma}
\newtheorem{claim}[theorem]{Claim}
\newtheorem{corollary}[theorem]{Corollary}
\newtheorem{proposition}[theorem]{Proposition}
\theoremstyle{definition}
\newtheorem{definition}[theorem]{Definition}

\newcommand{\cqed}{\renewcommand{\qedsymbol}{$\lrcorner$}}

\begin{document}

  \date{}

  \author{
  Sylvain Guillemot
  \thanks{
	Institute for Computer Science and Control, Hungarian Academy of Sciences (MTA SZTAKI)
	\texttt{sguillem@sztaki.hu}.}
  \and
  D\'{a}niel Marx
  \thanks{
	Institute for Computer Science and Control, Hungarian Academy of Sciences (MTA SZTAKI)
	\texttt{dmarx@cs.bme.hu}.}
  }

  \title{Finding small patterns in permutations in linear time}

\begin{titlepage}
\def\thepage{}
\thispagestyle{empty}
\maketitle
\begin{abstract}
  Given two permutations $\sigma$ and $\pi$, the \textsc{Permutation
    Pattern} problem asks if $\sigma$ is a subpattern of $\pi$. We
  show that the problem can be solved in time $2^{O(\ell^2\log \ell)}\cdot
  n$, where $\ell=|\sigma|$ and $n=|\pi|$. In other words, the problem is
  fixed-parameter tractable parameterized by the size of the
  subpattern to be found.

  We introduce a novel type of decompositions for permutations and a
  corresponding width measure. We present a linear-time algorithm that
  either finds $\sigma$ as a subpattern of $\pi$, or finds a
  decomposition of $\pi$ whose width is bounded by a function of
  $|\sigma|$. Then we show how to solve the \textsc{Permutation
    Pattern} problem in linear time if a bounded-width decomposition
  is given in the input.
\end{abstract}
\end{titlepage}

\section{Introduction}

A permutation of length $n$ is a bijective mapping $\pi:[n]\to [n]$;
one way to represent it is as the sequence of numbers
$\pi(1)\pi(2)\dots\pi(n)$. We say that a permutation $\pi$
written in this notation {\em contains} permutation $\sigma$ if $\pi$
has a (not necessarily consecutive) subsequence where the relative
ordering of the elements is the same as in $\sigma$. In this case, we
say that $\sigma$ is a {\em subpattern} of $\pi$; otherwise, $\pi$
{\em avoids} $\sigma$. For example, 3215674 contains the pattern 132,
since the subsequence 154 is ordered the same way as 132. On the other
hand, the permutation avoids 4321: it does not contain a descending
subsequence of 4 elements.

Counting the number of permutations avoiding a fixed pattern $\sigma$
has been a very actively investigated topic of enumerative
combinatorics. It was shown that for every length $n$, the number
of permutations avoiding the pattern 123 and the number of
permutations avoiding the pattern 231 are the same, namely the $n$th
Catalan number
\cite{DBLP:books/aw/Knuth68,MacMahon,MR829358,MR2465405}, which is
asymptotically $4^{n+o(1)}$. Around 1990, Stanley and Wilf conjectured
that for every fixed pattern $\sigma$, the number of permutations of
length $n$ avoiding $\sigma$ can be bounded by $c^n$ for some constant
$c$ depending on $\sigma$ (whereas the total number of permutations is
$n!=2^{\Theta(n\log n)}$). This conjecture has been proved by Marcus
  and Tardos \cite{DBLP:journals/jct/MarcusT04} in 2004.

  The algorithmic study of permutations avoiding fixed patterns was
  motivated first by the observation that permutations sortable by
  stacks and deques can be characterized by certain forbidden patterns
  and such permutations can be recognized in linear time
  \cite{DBLP:books/aw/Knuth68,DBLP:conf/stoc/Pratt73,DBLP:journals/jal/RosentiehlT84}.
  In the \textsc{Permutation Pattern} problem, two permutations
  $\sigma$ and $\pi$ are given and the task is to decide if $\pi$
  contains $\sigma$. In general, the problem is $\np$-hard
  \cite{DBLP:journals/ipl/BoseBL98}. There are known polynomial-time
  solvable special cases of the problem: for example, when $\sigma$ is
  the identity permutation $12\cdots k$, then the problem is a special
  case of \textsc{Longest Increasing Subsequence}, whose
  polynomial-time solvability is a standard textbook exercise \cite{DBLP:books/daglib/0023376}.  Other
  polynomial cases include the cases when $\sigma$ and $\pi$ are
  separable \cite{DBLP:journals/ipl/BoseBL98}, or when both $\sigma$
  and $\pi$ avoids 321 \cite{DBLP:conf/isaac/GuillemotV09}. For more
  background, the reader is referred to the survey of Bruner and
  Lackner \cite{DBLP:journals/corr/abs-1301-0340}.

  The \textsc{Permutation Pattern} problem can be solved by brute
  force in time $O(n^\ell)$, where $\ell=|\sigma|$ and $n=|\pi|$. This has
  been improved to $O(n^{0.47\ell+o(\ell)})$ by Ahal and Rabinovich
  \cite{DBLP:journals/siamdm/AhalR08}. These results imply that the
  problem is polynomial-time solvable for fixed pattern $\sigma$, but
  as the size of $\sigma$ appears in the exponent of the running time,
  this fact is mostly of theoretical interest only.  Our main result
  is an algorithm where the running time is linear for fixed $\sigma$
  and the size of $\sigma$ appears only in the multiplicative factor
  of the running time.
\begin{theorem}\label{th:main}
  \textsc{Permutation Pattern} can be solved in time $2^{O(\ell^2\log
    \ell)}\cdot n$, where $\ell=|\sigma|$ is the length of the pattern and
  $n=|\pi|$.
\end{theorem}
In other words, \textsc{Permutation Pattern} is fixed-parameter
tractable parameterized by the size of the pattern: recall that a
problem is fixed-parameter tractable with a parameter $\ell$ if it can be
solved in time $f(\ell)\cdot n^{O(1)}$, where $f$ is an arbitrary
computable function depending only on $\ell$; see
\cite{MR2001b:68042,grohe-flum-param}. The fixed-parameter
tractability of \textsc{Permutation Pattern} has been an open question
implicit in previous work.

\begin{figure*}
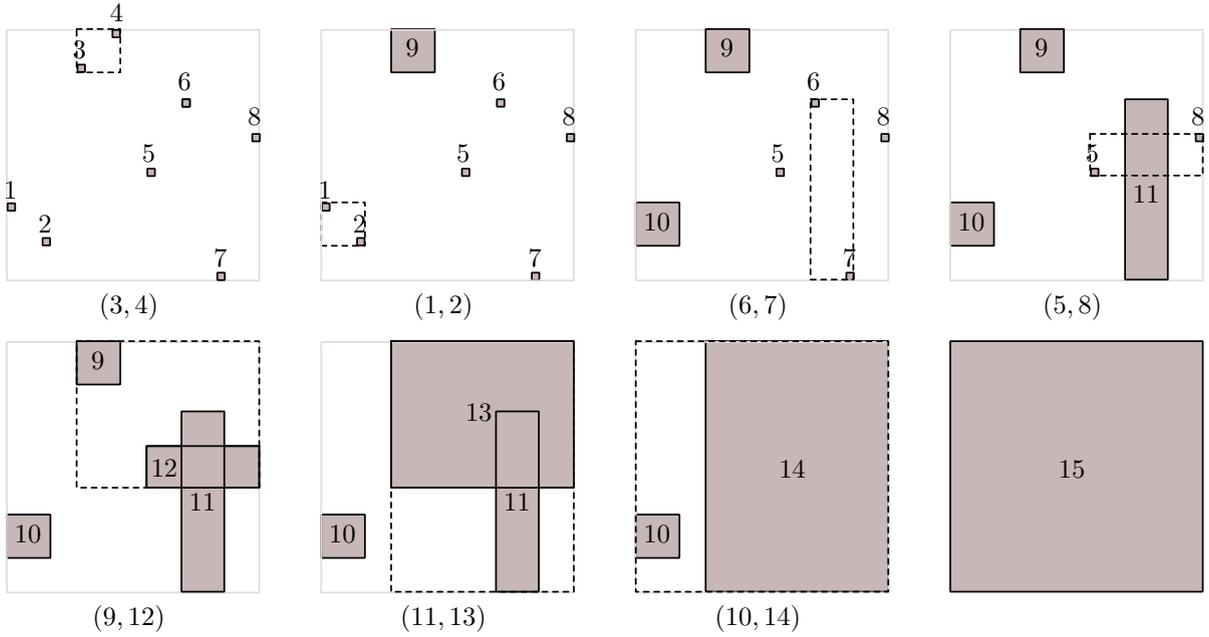

\begin{center}
{\small \svg{0.99\linewidth}{merge}}
\caption{A possible decomposition of the permutation 32784615. The
  dashed rectangles and the pairs of numbers below the figures show
  the next two rectangles to be merged. We follow the convention that
  the rectangle created in the $i$-th merge is labeled
  $n+i$.}\label{fig:merge}
\end{center}
\end{figure*}

The main technical concept in the proof of Theorem~\ref{th:main} is a
novel form of decomposition for permutations. The decomposition can be
explained most intuitively using a geometric language. Given a
permutation $\pi$ of length $n$, one can represent it as the set of
points $(1,\pi(1))$, $(2,\pi(2))$, $\dots$, $(n,\pi(n))$ in the
2-dimensional plane. We can view these points as a family of
degenerate rectangles, each having width and height 0. Starting with
this family of $n$ degenerate rectangles, our decomposition consists
of a sequence of families of rectangles, where the next family is
created from the previous one by a {\em merge} operation. The merge
operation removes two rectangles $R_1$, $R_2$ from the family and
replaces them with their bounding box, that is, the smallest rectangle
containing both (see Figure~\ref{fig:merge}). The decomposition is a
sequence of $n-1$ merges that eventually replaces the whole family
with a single rectangle. Note that the rectangles created by the
merges are not necessarily disjoint. We define a notion of {\em width}
for a family of rectangles, which roughly corresponds to the maximum
number of other rectangles a rectangle can ``see'' either horizontally
or vertically. The decomposition has width at most $d$ if the
rectangle family has width at most $d$ at every step of the
decomposition. Let us observe that the merge operation can increase
the width (by creating a large rectangle that sees many other
rectangles) or it can decrease width (since if a rectangle sees both of
the merged rectangles, then it sees one less rectangle after the merge
operation). Therefore, whether it is possible to maintain bounded
width during a sequence of merge operations is a very subtle and
highly nontrivial question.

The proof of Theorem~\ref{th:main} follows from the following two
results on bounded-width decompositions:
\begin{itemize}
\item[(1)] For every fixed pattern $\sigma$, there is a linear-time
  algorithm that, given a permutation $\pi$, either shows that
  $\sigma$ appears in $\pi$ or outputs a bounded-width decomposition
  of $\pi$ (Theorem~\ref{t:algo-decomp}).
\item[(2)] \textsc{Permutation Pattern} can be solved in linear time
  if a bounded-width decomposition for $\pi$ is given in the input (Theorem~\ref{th:dp}).
\end{itemize}
The proof of (1) needs to show how to find the next mergeable pair in
the decomposition. We argue that there have to be two rectangles that
are ``close'' in a certain sense (ensuring that the width is still
bounded after the merge), otherwise the rectangles are so much spread
out that a result of Marcus and Tardos
\cite{DBLP:journals/jct/MarcusT04} guarantees that every permutation
of length $\ell$ (in particular, $\sigma$) appears in $\pi$. The
implementation of this idea needs careful control of the global
structure of the rectangles. Therefore, the algorithm is not based on
simply merging pairs in a greedy way: instead of showing that a
mergeable pair always exist, what we show is that if the global property
holds, then it is possible to merge two rectangles such that the
global property still holds after the merge.

The algorithm in (2) uses dynamic programming the following way. Given
a family of rectangles, we can define a {\em visibility graph} where the
vertex set is the set of rectangles and two rectangles are adjacent if
their horizontal or vertical projections intersect. For our purposes,
the meaning of this graph is that if two rectangles $R_1$ and $R_2$
are nonadjacent, then the relative position of $x\in R_1$ and $y\in
R_2$ follows from the relative position of $R_1$ and $R_2$; on the
other hand, if $R_1$ and $R_2$ are adjacent, then the relative
position of $x$ and $y$ can depend on exactly where they appear in
$R_1$ and $R_2$. The fact that the decomposition has bounded width
implies that the visibility graph of the rectangle family at every step
has bounded degree. We enumerate every set $K$ of at most $\ell$
rectangles that induces a connected graph in the visibility graph; as
the visibility graph has bounded degree, the number of such sets is
linear in $n$. The subproblems of the dynamic programming are as
follows: at each step of the decomposition, for each set $K$ of size
at most $\ell$ that is connected in the visibility graph, we have to
enumerate every pattern that appears in the points contained in the
rectangles (and the possible distribution of the elements of the
pattern among the rectangles). In each step of the decomposition, only
those subproblems have to be updated that involve the merged
rectangles, and this can be done efficiently using the information at
hand. At the last step, there is only a single rectangle containing
every element of $\pi$; thus the subproblems for this single rectangle
tell us if $\sigma$ is contained in $\pi$.

The fixed-parameter tractability of the \textsc{Permutation Pattern}
problem seems to be very fragile: every reasonable extension or
generalization of the problem (e.g., introducing colors or introducing
additional constraints such as certain elements of the pattern having to
appear consecutively) turned out to be $\wone$-hard
\cite{DBLP:journals/corr/abs-1301-0340,DBLP:conf/isaac/GuillemotV09,
DBLP:conf/isaac/Guillemot11}. In Section~\ref{sec:hardness-results},
we prove the $\wone$-hardness of another colored variant of
the problem and then infer that the natural 3-dimensional
generalization of \textsc{Permutation Pattern} is also $\wone$-hard.

The reason for the $2^{O(\ell^2\log \ell)}$ dependence on $\ell$ in
Theorem~\ref{th:main} is the following. Recall that Marcus and
Tardos~\cite{DBLP:journals/jct/MarcusT04} proved that for every fixed
permutation $\sigma$, there is a constant $c$ such that the number of
permutations of length $n$ that avoid $\sigma$ can be bounded by
$c^{n}$. In their proof, the constant $c$ is exponential in the length
of $\sigma$, but it might be true that the result holds with
polynomially bounded $c$. Our algorithm for finding a decomposition
relies on the proof of Marcus and Tardos, and the bound we get on the
width is exponential in the length $\ell$ of $\sigma$, which implies that
the algorithm using this decomposition has running time $2^{O(\ell^2\log
  \ell)}\cdot n$. Improving the exponential bound in the result of Marcus
and Tardos to a polynomial would immediately imply that we can find a
decomposition of polynomially bounded width, and then the running time
would be $2^{O(\ell\log \ell)}\cdot n$.

We investigate a specific class of permutations for which we can give
improved bounds. A sequence is {\em monotone} if it is either
increasing or decreasing; we say that a permutation is {\em $t$-monotone}
if it can be partitioned into $t$ monotone (not necessarily
consecutive) subsequences. For the special case when $\pi$ is
$t$-monotone we show that it is possible to find a decomposition of width
polynomially bounded by $t$. Moreover, we show that there is a
very simple way of solving \textsc{Permutation Pattern} when $\pi$ is
$t$-monotone. The crucial observation is that if, given $t$ monotone
sequences, the task is to select some elements from each sequence such
that they form a specific pattern, then this can be encoded as a
constraint satisfaction problem (CSP) having a majority
polymorphism. It is well-known that CSPs with a majority polymorphism
is polynomial-time solvable, which gives us a polynomial-time solution
for \textsc{Permutation Pattern} for a specific distribution of
the elements of $\sigma$ among the $t$ monotone subsequences of $\pi$.
Finally, we can try all possible way of distributing the elements of the
pattern among the monotone sequences, yielding a very compact proof
for the fixed-parameter tractability of \textsc{Permutation Pattern}
on $t$-monotone permutations.

On a high level, our algorithm can be described by the following
scheme: either the permutation $\pi$ has no bounded-width decomposition,
in which case we can answer the problem immediately, or we can find a
bounded-width decomposition, in which case we can use an algorithm working
on the decomposition. This win/win scenario is very similar to how the
notion of treewidth is used for many graph problems: high treewidth
implies an immediate answer and bounded-treewidth graphs can be
handled using standard techniques. This idea was used, for example, in the classical
work of Bodlaender
\cite{DBLP:journals/ijfcs/Bodlaender94,DBLP:journals/jal/Bodlaender93}
and more recently in the framework of bidimensionality for planar
graphs \cite{DBLP:journals/jacm/DemaineFHT05}. However, it has to be
pointed out that our notion of decomposition is very different from
tree decompositions. The main property of tree decompositions is that
the graph is broken down into parts that interact with each other only
via a small boundary. Nothing similar happens in our decomposition:
when we merge two rectangles, then the points appearing in the two
rectangles can have very complicated relations. Perhaps it is even
misleading to call our notion a ``decomposition'': it would be more
properly described as a construction scheme that maintains a notion of
bounded-degreeness throughout the process. It would be interesting to
see if there is a corresponding graph-theoretic analog for this
scheme, which might be useful for solving some graph-theoretical
problem.

\iffull The paper is organized as follows. Section~\ref{sec:definitions} introduces notation, including a somewhat nonstandard way of looking at permutations as a labeled point set. Section~\ref{sec:decompositions} defines our notion of decomposition and width measure, and observes some properties. Section~\ref{sec:grid-obstructions} presents our algorithm for finding a decomposition. Section~\ref{sec:solving-subpattern} shows how to solve the \textsc{Permutation Pattern} problem given a decomposition. Section~\ref{sec:hardness-results} proves hardness results for some natural generalizations of the problem. Section~\ref{sec:t-monotone} investigates the special case when $\pi$ is a $t$-monotone permutation.
\fi


\section{Definitions}\label{sec:definitions}

A {\em permutation} of length $n$ is a bijection $\pi:[n]\to [n]$. It will be convenient for us to look at permutations from a more geometric viewpoint by considering them as point sets, as our decomposition can be explained conveniently in terms of families of points and rectangles.
 A \emph{point} is an element $p = (x,y) \in \nat^2$; we denote $\pr_1(p) = x$ and $\pr_2(p) = y$ (these are called the $x$-coordinate and $y$-coordinate of $p$). A \emph{point set} is a finite set of points; it is in \emph{general position} if no two points have the same $x$-coordinate or the same $y$-coordinate. We define a \emph{permutation} as a pair $\pi = (S,P)$, where $S$ is a subset of positive integers and $P : S \rightarrow \nat^2$ is an injection such that $P(S)$ is a point set in general position. For a permutation $\pi=(S,P)$, we use $S(\pi)$ to refer to the set $S$, and define the \emph{length} of $\pi$ as $|\pi| = |S(\pi)|$. Given $S' \subseteq S$, we define the permutation $\pi | S' = (S', P | S')$.

Let us discuss how permutations are represented in algorithms. 
 We say that a permutation $\pi = (S,P)$ of length $n$ is \emph{reduced} if $S = [n]$ and $P(S) \subseteq [n] \times [n]$. 
A reduced permutation can be represented naturally as an array of $n$ points in $[n]\times [n]$. We require that the permutation given as an input of an algorithm is reduced and has this representation; we mainly use this  assumption to ensure that we can sort the points by $x$- or $y$-coordinate in linear time. 
Note that if we consider a permutation to be a bijection $\pi:[n]\to [n]$, then it is straightforward to obtain such a representation.

Given $p,p' \in S$ and $\alpha \in \{1,2\}$, we denote $p <^{\pi}_{\alpha} p'$ iff $\pr_{\alpha}(P(p)) < \pr_{\alpha}(P(p'))$. Given two permutations $\sigma$ and $\pi$, a mapping $\phi : S(\sigma) \rightarrow S(\pi)$ is an \emph{embedding of $\sigma$ into $\pi$} iff for every $p,p' \in S(\sigma)$, for each $\alpha \in \{1,2\}$, $p <^{\sigma}_{\alpha} p'$ iff $\phi(p) <^{\pi}_{\alpha} \phi(p')$. We say that $\sigma$ is a \emph{subpattern} of $\pi$, or that $\pi$ \emph{contains} $\sigma$, if there is an embedding of $\sigma$ into $\pi$. (Intuitively, we can represent a permutation as a 0-1 matrix where every column and every row contains at most one cell with 1 in it; then $\sigma$ is a subpattern of $\pi$ if it corresponds to a submatrix of the matrix representation of $\pi$).
We define the following decision problem:

\prob{\textsc{Permutation Pattern}}{Two reduced permutations $\sigma$ and $\pi$.}{Is $\sigma$ a subpattern of $\pi$?}
\medskip

For a given instance $(\sigma,\pi)$ of the problem, we will denote $\ell = |\sigma|$ and
$n = |\pi|$. Besides points and sets of points, we will be dealing with rectangles
and sets of rectangles as well.  Given two positive integers $p,q$ with $p \leq
q$, we define the \emph{interval} $[p,q] = \{p,p+1,\ldots,q\}$; note
that we only consider discrete intervals. Given two intervals $I =
[p,q]$ and $I' = [p',q']$, we denote $I < I'$ iff $q < p'$. A
(axis-parallel) \emph{rectangle} is a set $R = I \times J$ where $I,J$
are two intervals; we denote $I_1(R) = I$ and $I_2(R) = J$. 


\section{Decompositions}\label{sec:decompositions}

The purpose of this section is to introduce the decomposition used by
the main algorithm and observe some of its properties. A {\em
  rectangle family} is a set of rectangles indexed by a subset of
natural numbers; formally, a rectangle family is a pair $\R = (S,R)$,
where $S\subseteq \mathbb{N}$ is a set and $R$ maps each element $i
\in S$ to a rectangle $R(i)$. For a rectangle family $\R = (S,R)$, we use
$S(\R)$ to refer to the set $S$ and we define the size of $\R$ as $|\R| = |S(\R)|$.
Note that a point is a degenerate rectangle, and thus a permutation can also be
viewed as a rectangle family. We define the operation of
{\em merging} two rectangles in a family as follows. Given two elements
$i,j \in S$ and $k \notin S$, we denote by $\R[i,j \rightarrow k]$ the rectangle
family $\R' = (S',R')$ where $S' = S - \{i,j\} + \{k\}$, $R'(p) = R(p)$ for every
$p \in S- \{i,j\}$, and $R'(k)$ is the smallest rectangle enclosing $R(i) \cup
R(j)$. That is, we replace rectangles $R(i)$ and $R(j)$ by their bounding
box, and assign the index $k$ to the new rectangle.

Our notion of decomposition is defined as follows.
\begin{definition}
  Let $\pi = (S,P)$ be a permutation of length $n$. A \emph{decomposition} of
  $\pi$ is a sequence $\D = (\R_0,\ldots,\R_s)$ of rectangle families such that:
  \begin{itemize}[noitemsep]
\item[(i)] $\R_0 = \pi$;
\item[(ii)] there exists a sequence of integers $ k_1 < k_2 < \ldots < k_s$
such that $\max S<k_1$ and for every $1 \leq p \leq s$, there exist $i,j \in S(\R_{p-1})$  such that
$\R_p = \R_{p-1}[i,j \rightarrow k_p]$;
\item[(iii)] $|\R_s| = 1$.
\end{itemize}
\end{definition}
That is, in each step we are merging two rectangles to create a new
rectangle. Observe that by Point (iii) we have $s = n-1$, i.e. the decomposition
contains $n$ rectangle families. This means that the obvious
representation of the decomposition can have size
$\Omega(n^2)$. However, let us observe that it is sufficient to list
the pairs of rectangles that are merged in each step. Therefore, we
can compactly represent the decomposition in space $O(n)$ by the
{\em merge sequence} $\Sigma = \sigma_1 \ldots \sigma_s$, where
for each $1 \leq p \leq s$ we have $\sigma_p = (i,j,k_p)$ if $\R_p =
\R_{p-1} [i,j \rightarrow k_p]$.

Next we define a notion of width for permutations. For $\alpha \in \{1,2\}$,
we say that two rectangles $R,R'$ \emph{$\alpha$-view} each other if
$I_{\alpha}(R)$ intersects $I_{\alpha}(R')$. Let $\R = (S,R)$ be a rectangle
family. Given $i \in S$ and $\alpha \in \{1,2\}$, we define $\view_{\alpha}(\R,i)$
as the set of elements $j \in S - \{i\}$ such that $R(i)$ and $R(j)$ $\alpha$-view
each other. Given $i \in S$, we define $\view(\R,i) = \max_{\alpha \in \{1,2\}}
|\view_{\alpha}(\R,i)|$. Note that we define the number $\view(\R,i)$ as the
maximum of the two cardinalities rather as the cardinality of the union, for
reasons that will become clear later. Let $d$ be an integer. We say that a rectangle family
$\R$ is \emph{$d$-wide} if $\view(\R,i) < d$ holds for every $i \in S(\R)$. We
say that a decomposition $\D = (\R_0,\ldots,\R_s)$ of $\pi$ is $d$-wide if
each rectangle family $\R_p$ is $d$-wide. Observe that it is enough to ask
whenever $\R_p$ merges rectangles $i$ and $j$ to produce rectangle $k$,
then $\view(\R_p,k) < d$: indeed, the view number of a rectangle can increase
only if it views $k$ but not $i,j$, in which case it is upper bounded by the view
number of $k$. We define the {\em width} of a permutation $\pi$,
denoted by $w(\pi)$, as the minimum $d$ such that $\pi$ has a $d$-wide
decomposition.

\subsection{Basic properties}\label{sec:props}

We observe that width is monotone for subpatterns:
\begin{lemma}\label{lem:subwidth}
If $\sigma$ is a subpattern of $\pi$, then $w(\sigma)\le w(\pi)$.
\end{lemma}
\iffull
\begin{proof}
  It is sufficient to show that if $S'$ is a subset of $S(\pi)$, then
  $w(\pi|S')\le w(\pi)$; in fact, by induction, it is sufficient to show this for the case when $S'=S-
\{j_1\}$ for some $j_1\in S(\pi)$.  
Consider a $d$-wide decomposition $\D=(\R_0,\dots, \R_{n-1})$ of $\pi$. We
modify the decomposition as follows. There is a unique step $i_1$ in $\D$
when $j_1$ is merged with some rectangle $j_2$ and they are replaced by the
bounding box $j_3$. If this is the last step of the decomposition,
then it is clear that removing this step and removing $j_1$ from each
of $\R_0$, $\dots$, $\R_{n-2}$ results in a $d$-wide decomposition of
$\pi|S'$.

Otherwise, suppose that this is not the last step. Then there is a
unique step $i_2>i_1$ when $j_3$ is merged with some rectangle $j_4$. We
remove step $i_1$ and modify step $i_2$ such that $j_4$ is merged with
$j_2$ instead of $j_3$. Therefore, we obtain a decomposition $\D'=(\R'_0,\dots, \R'_{n-2})$, where
\begin{itemize}[noitemsep]
\item for $0 \le i < i_1$, rectangle family $\R'_i$ is obtained from $\R_i$ by removing element $j_1$.
\item for $i_1 \le i \le i_2-2$, rectangle family $\R'_i$ is obtained from $\R_{i+1}$ by replacing $j_3$ with $j_2$,
\item for $i_2-1 \le i\le n-2$, rectangle family $\R'_i$ is obtained from  $\R_{i+1}$ by modifying only a single rectangle, namely the one whose construction involved $j_1$.
\end{itemize}
In the last case, the single modified rectangle cannot become larger:
it is constructed as the merge of one fewer points than in
$\D$. Therefore, in all cases, the fact that every $\R_i$ is $d$-wide
implies that every $\R'_i$ is $d$-wide. Thus $\D'$ is a $d$-wide
decomposition of $\pi|S'$ and $w(\pi|S')\le w(\pi)$ follows.
\end{proof}
\fi

Next, we observe a relation between the width and the
existence of close pairs of points. Let $\pi$ be a permutation. Given
$p,p' \in S(\pi)$ and $\alpha \in \{1,2\}$, if $p <^{\pi}_{\alpha} p'$
then we denote $\Int_{\alpha}(\pi,p,p') = \{ p'' \in S(\pi) \mid p <^{\pi}_{\alpha}
p'' <^{\pi}_{\alpha} p' \}$; if $p' <^{\pi}_{\alpha} p$ then we let $\Int_{\alpha}(\pi,p,p')
:= \Int_{\alpha}(\pi,p',p)$. For an integer $d$, we say that $\{p,p'\}
\subseteq S(\pi)$ is a \emph{$d$-close pair} of $\pi$ if for each $\alpha \in \{1,2\}$
it holds that $|\Int_{\alpha}(\pi,p,p')| < d$. Let us observe that the existence
of a $d$-close pair is a necessary condition for having a $d$-wide
decomposition: the first pair $\{j_1,j_2\}$ of points merged in the
decomposition should be $d$-close: otherwise the rectangle family
obtained by replacing $j_1$ and $j_2$ with their bounding box would
have a view number greater than $d$.

\begin{proposition}\label{p:close}
If $w(\pi)\le d$, then $\pi$ has a $d$-close pair.
\end{proposition}
\iffull Note that by Lemma~\ref{lem:subwidth}, in fact every subpermutation of
$\pi$ has a $d$-close pair. As we shall see in
Section~\ref{sec:grid-obstructions}, the existence of $d$-close pairs
in the subpermutations approximately characterizes the width of the
permutation.
\fi

\iffull
\subsection{Separable permutations}

In this section, we relate our width measure to the well-known notion of separable
permutations (note that this connection is not needed for the main algorithmic results of the paper). 
The separable permutations are the permutations that are totally decomposable under the 
\emph{substitution decomposition} \cite{MR84,AA05}, and we show in Proposition \ref{p:separable}
below that they correspond to permutations of width at most 1.

We first define the operation of \emph{substitution} for permutations. Let $\pi = (S,P)$ and
$\pi' = (S',P')$ be two permutations with $S \cap S' = \emptyset$. Given $x \in S$, we define
the permutation $\pi [x \leftarrow \pi']$ as follows. This is a permutation
$\pi'' = (S'',P'')$, where $S'' = S - \{x\}  + S'$, and such that two elements
$p,p' \in S''$ have the following relations: (i) if $p,p' \in S$ then
$p <^{\pi''}_{\alpha} p'$ iff $p <^{\pi}_{\alpha} p'$; (ii) if $p,p' \in S'$, then
$p <^{\pi''}_{\alpha} p'$ iff $p <^{\pi'}_{\alpha} p'$; (iii) if $p \in S, p' \in S'$, then
$p <^{\pi''}_{\alpha} p'$ iff $p <^{\pi}_{\alpha} x$. 

\begin{proposition} \label{p:subst} Given two permutations $\pi$ and $\pi'$,
and given $x \in S(\pi)$, it holds that $w(\pi[x \leftarrow \pi']) = \max(w(\pi),w(\pi'))$.
\end{proposition}

\begin{proof} Let $d = w(\pi)$ and $d' = w(\pi')$, and let $\pi'' = \pi[x \leftarrow \pi']$.
As $\pi$ and $\pi'$ are subpatterns of $\pi''$, it follows that $w(\pi'') \geq \max(d,d')$
by Lemma \ref{lem:subwidth}. Let us show that $w(\pi'') \leq \max(d,d')$. 
Let $\D = (\R_0,\ldots,\R_r)$ be a $d$-wide decomposition of $\pi$ and let $\D' =
(\R'_0,\ldots,\R'_s)$ be a $d'$-wide decomposition of $\pi'$. We assume w.l.o.g. that
$\D'$ produces a sequence of indices $k'_1 < \ldots < k'_s$ and $\D$ produces a sequence
of indices $k_1 < \ldots < k_r$ such that $\max S(\pi'') < k'_1$ and $k'_s < k_1$. We construct a
decomposition $\D'' = (\R''_0,\ldots,\R''_t)$ of $\pi''$ as follows. We first simulate
the merges of $\D'$, then once the points of $\pi'$ have been merged into a single
rectangle we simulate the merges of $\D$. More precisely, we start with $\R''_0 = \pi''$, and:
\begin{itemize}
\item for $1 \leq p \leq s$, if $\R'_p = \R'_{p-1}[i,j \rightarrow k]$ then $\R''_p = \R''_{p-1}[i,j \rightarrow k]$;
\item for $1 \leq p \leq r$, if $\R_p = \R_{p-1}[i,j \rightarrow k]$ then $\R''_{s+p} = \R''_{s+p-1}[i',j' \rightarrow k]$,
where $i',j'$ are obtained from $i,j$ by replacing $x$ with $k'_s$.
\end{itemize}
Observe that a rectangle created in the first step views the same rectangles as in $\D'$, while
a rectangle created in the second step views the same rectangles as in $\D$. We conclude
that $\D''$ is a $d''$-wide decomposition of $\pi''$ with $d' = \max(d,d')$.
\end{proof}

We recall that the separable permutations can be defined as follows \cite{DBLP:journals/ipl/BoseBL98}.
A permutation $\pi$ is \emph{increasing} (resp. \emph{decreasing}) if for each $p,p' \in S(\pi)$ it holds
that $p <^{\pi}_1 p'$ iff $p <^{\pi}_2 p'$ (resp. $p' <^{\pi}_2 p$). A permutation $\pi$ is \emph{monotone} if
it is increasing or decreasing. The \emph{separable} permutations is the smallest
class of permutations that contains the monotone permutations and is closed under
substitution; alternatively, they are the permutations that do not contain $2~4~1~3$
or $3~1~4~2$.

\begin{proposition} \label{p:separable} A permutation $\pi$ is separable iff $w(\pi) \leq 1$.
\end{proposition}

\begin{proof} As the monotone permutations have width at most $1$, it follows from
Proposition \ref{p:subst} that the separable permutations have width at most 1.
Conversely, if a permutation $\pi$ is not separable, then $\pi$ contains $2~4~1~3$
or $3~1~4~2$; as these two permutations have no 1-close pair,
they have width at least 2 by Proposition \ref{p:close}, which implies that
$w(\pi) \geq 2$ by Lemma \ref{lem:subwidth}.
\end{proof}
\fi

\subsection{Grids}\label{sec:grids}

In this section, we define certain permutations with a grid-like structure, and we
characterize their widths. The main interest of these permutations is that they
serve as obstruction patterns to small width; moreover, we will see in Section
\ref{sec:grid-obstructions} that they are the only obstructions in an approximate
sense.

Given an interval $I$, we say that a sequence $P = (I_1,\ldots,I_s)$
of intervals is a \emph{partition} of $I$ if (i) the $I_j$'s are
disjoint and their union is $I$, and (ii) $I_1 < I_2 < \ldots < I_s$.
Consider the rectangle $R = I \times J$, and fix two integers
$r,s$. An \emph{$r \times s$-gridding} of $R$ is a pair $G =
(P_1,P_2)$, where $P_1 = (I_1,\ldots,I_r)$ is a partition of $I$, and
$P_2 = (J_1,\ldots,J_s)$ is a partition of $J$. Fix $x \in [r], y \in [s]$.
We call $I_x$ the \emph{$x$th column} of $G$, and $J_y$ the
\emph{$y$th row} of $G$; the rectangle $G(x,y) := I_x \times J_y$
is called the \emph{$(x,y)$th-cell} of $G$. If $M$ is a point set, we
say that $M$ \emph{contains an $r \times s$-grid} if there exists an
$r \times s$-gridding $G$ such that for every $x \in [r], y \in [s]$,
$G(x,y)$ intersects $M$. By extension, if $\pi = (S,P)$ is a permutation,
we say that $\pi$ \emph{contains an $r \times s$-grid} if $P(S)$ does.

\begin{figure}
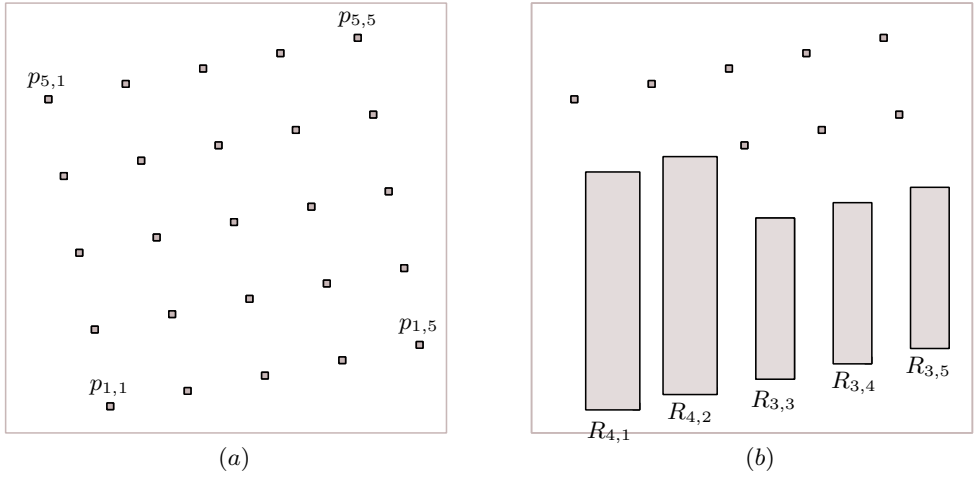

\begin{center}
{\footnotesize \svg{0.8\linewidth}{grid}}
\end{center}
\caption{(a) A $5\times 5$ canonical grid. (b) A step of the decomposition in the proof of Proposition~\ref{p:canongrid}.}
\label{fig:grid}
\end{figure}

An \emph{$r \times s$-grid permutation} is a permutation of length $r s$
that contains an $r \times s$-grid. Observe that a permutation contains an
$r \times s$-grid if and only if it contains an $r \times s$-grid permutation.
Furthermore, observe that if a permutation contains an $r \times r$-grid,
then it contains every permutation of length $r$; this fact will be crucial for
our algorithm. The \emph{canonical $r \times s$-grid permutation} is the
permutation $\pi$ corresponding to the point set $\{((j-1) s + (s-i+1), (i-1) r + j)| 1 \le i \le s, 1 \le j \le r\}$;
let us denote by $p_{i,j}$ the element of  $S(\pi)$ corresponding to point $((j-1) s + (s-i+1), (i-1) r + j)$.
Intuitively, $p_{i,j}$ is the point in row $i$ and column $j$, where rows are numbered
from bottom to top and columns are numbered from left to right (see Figure~\ref{fig:grid}(a)).
Note that the indexing of points $p_{i,j}$ departs from the convention used for points in
cartesian coordinates, i.e. point $p_{i,j}$ is inside the $(j,i)$th cell of the gridding of $\pi$.

The following result shows that $r \times r$-grid permutations have width
$\Omega(r)$.

\begin{proposition}\label{p:gridwidth}
If $\pi$ is a $(2r+4) \times (2r+4)$-grid permutation, then $w(\pi) \geq r$.
\end{proposition}
\iffull
\begin{proof}
  Consider a decomposition $(\R_0,\ldots,\R_{s})$ of $\pi$.  Let
  $\R_t$ be the first family in this sequence that includes a
  rectangle $R$ containing points from two nonadjacent rows or from
  two nonadjacent columns. Suppose without loss of generality that $R$
  contains points from rows $y_1$ and $y_2$ with $y_2-y_1>1$. Consider
  the set $X$ of $2r+4$ points of $\pi$ in row $y_1+1$. In family
  $\R_{t-1}$, no rectangle contains points from two nonadjacent
  columns, thus at most two points of $X$ can be contained in each
  rectangle of $\R_{t-1}$, i.e., points of $X$ are contained in at
  least $r+2$ rectangles. At most two of these rectangles can
  participate in the merge that created rectangle $R$ in
  $\R_{t}$. Therefore, at least $r$ of these rectangles survive in
  $\R_t$ and are distinct from $R$. All of these rectangles 2-view
  $R$, hence $\R_t$ (and therefore the decomposition) cannot be
  $r$-wide.
\end{proof}
\fi

\begin{proposition}\label{p:canongrid}
If $\pi$ is the canonical $r \times r$-grid permutation, then $w(\pi) = r$.
\end{proposition}
\iffull
\begin{proof}
  We first show that $w(\pi) \geq r$. Consider
  two distinct elements $p = p_{i,j}$ and $p' = p_{i',j'}$ in $S(\pi)$. We
  have $|\Int_1(\pi,p,p')| = | (j'-j) r + (i-i') | - 1$ and
  $|\Int_2(\pi,p,p')| = | (i'-i) r + (j'-j) | - 1$. Observe that if
  $|j'-j| \geq 2$ then $|\Int_1(\pi,p,p')| \geq r-1$, and likewise if
  $|i'-i| \geq 2$ then $|\Int_2(\pi,p,p')| \geq r-1$. Suppose that
  $\{p,p'\}$ is a $(r-1)$-close pair of $\pi$. We then have $0 \leq
  |i'-i| \leq 1$ and $0 \leq |j'-j| \leq 1$, and one of them is equal
  to 1; we suppose w.l.o.g. that $i'-i = 1$. Then $|\Int_2(\pi,p,p')| <
  r-1$ implies that $j-j' = 1$, and thus $|\Int_1(\pi,p,p')| = r >
  r-1$, contradiction. It follows that $\pi$ has no $(r-1)$-close
  pair, and thus $w(\pi) \geq r$ by Proposition~\ref{p:close}.

  For $1\le j \le r$, let $R_{1,j}$ be the rectangle containing only point
  $p_{1,j}$. We define a decomposition that first merges $R_{1,1}$ and
  $p_{2,1}$ to obtain $R_{2,1}$; then $R_{1,2}$ and $p_{2,2}$ to obtain
  $R_{2,2}$; $\dots$; then $R_{1,r}$ and $p_{2,r}$ to obtain $R_{2,r}$.
  We continue in a similar way with the next row: we merge $R_{2,1}$ and $p_{3,1}$ to
  obtain $R_{3,1}$; then $R_{2,2}$ and $p_{3,2}$ to obtain $R_{3,2}$;
  $\dots$; then $R_{2,r}$ and $p_{3,r}$ to obtain $R_{3,r}$ (see Figure~\ref{fig:grid}(b)). After repeating
  this process for each row, only $r$ rectangles $R_{r,1}$, $\dots$,
  $R_{r,r}$ remain. What needs
  to be observed is that when we merge $R_{i,j}$ and $p_{i+1,j}$ to obtain
  $R_{i+1,j}$, then $R_{i+1,j}$ 2-views only $R_{i+1,1}$, $\dots$,
  $R_{i+1,j-1}$, $R_{i,j+1}$, $\dots$, $R_{i,r}$ (i.e., $r-1$ rectangles) and does
  not 1-view any other rectangle. Therefore, the rectangle family
  is always $r$-wide. When there are only $r$ remaining
  rectangles, we can merge them in any order. We get an $r$-wide
  decomposition of $\pi$, showing that $w(\pi)\le r$.
\end{proof}

\fi

\iffull
\subsection{Tree representation}\label{sec:tree-representation}

Although it is not used explicitly in the paper, we can give an alternative representation of a decomposition by a labeled tree. We state in Proposition \ref{p:alt-charac} below a characterization of $d$-wide decompositions in terms of the associated tree.

A \emph{numbered tree} is a (directed) tree $T = (V,A)$, where (i) $V \subseteq \nat$, (ii) the leaves of $T$ precede the internal nodes in the natural ordering, (iii) for each arc $(i,j)$ it holds that $j < i$. We denote $L(T)$ the set of leaves of $T$, $I(T)$ the set of internal nodes of $T$ and $N(T)$ the set of nodes of $T$. Suppose that $\D = (\R_0,\ldots,\R_s)$ is a decomposition of a permutation $\pi$, we represent it by a binary numbered tree $T$ constructed as follows: (i) start with one vertex per element of $S(\pi)$; (ii) for $p$ going from $1$ to $s$, if $\R_{p} = \R_{p-1}[i,j \rightarrow k]$ then add a vertex $k$ with arcs $(k,i)$ and $(k,j)$. Conversely, if $T$ is a binary numbered tree with $L(T) = S(\pi)$, then there exists a decomposition $\D(\pi,T)$ whose associated tree is $T$.

We need the following additional definitions. Let $T$ be a numbered tree. Given two nodes $i,j$ of $T$, we denote $i <_T j$ (resp. $i \leq_T j$) if $j$ is a proper ancestor (resp. ancestor) of $i$. Fix a node $i \in N(T)$. We denote by $T(i)$ the subtree of $T$ rooted at $i$. We let $R(\pi,T,i)$ denote the bounding box of $\pi | L(T(i))$, and for each $\alpha \in \{1,2\}$ we let $I_{\alpha}(\pi,T,i) := I_{\alpha}(R(\pi,T,i))$. We let $S(T,i)$ denote the set of elements $j \in N(T)$ such that $j \leq i$ and that are maximal for $<_T$ with this property. We let $\R(\pi,T,i)$ denote the rectangle family $\R' = (S',R')$ with $S' = S(T,i)$ and for each $j \in S(T,i)$, $R(j) = R(\pi,T,j)$. Observe that if $\D(\pi,T) = (\R_0,\ldots,\R_s)$ and step $p$ produces index $j$, then $\R_p = \R(\pi,T,j)$. Finally, we define the \emph{restriction} of a numbered tree: if $T$ is a numbered tree and $X \subseteq L(T)$, then $T | X$ is the minimum homeomorphic subtree of $T$ containing the leaves of $X$.

\begin{proposition} \label{p:alt-charac} Let $\pi$ be a permutation, and let $T$ be a binary numbered tree with $L(T) = S(\pi)$. The following statements are equivalent:
\begin{itemize}
\item[(i)] $\D(\pi,T)$ is a $d$-wide decomposition of $\pi$;
\item[(ii)] for every $X \subseteq S(\pi)$ with $|X| \geq 2$, if $i$ is the minimum internal node of $T|X$ then $L(T|X(i))$ is a $d$-close pair of $\pi | X$.
\end{itemize}
\end{proposition}

\begin{proof} $(ii) \Rightarrow (i)$: We need to show that $\D(\pi,T)$ is a $d$-wide decomposition of $\pi$. Let $i \in I(T)$, and let $S = S(T,i)$ and $\R = \R(\pi,T,i)$. Fix $i' \in S$, we need to show that $\view(\R,i') < d$. Fix $\alpha \in \{1,2\}$. Consider $j \in S$ such that $I_{\alpha}(\pi,T,i') \subseteq I_{\alpha}(\pi,T,j)$ and $I_{\alpha}(\pi,T,j)$ is maximal with this property. As $\view_{\alpha}(\R,i') - \{j\} + \{i'\} \subseteq \view_{\alpha}(\R,j)$, we have $|\view_{\alpha}(\R,j)| \geq |\view_{\alpha}(\R,i')|$. We will thus show that $|\view_{\alpha}(\R,j)| < d$, which will imply that $|\view_{\alpha}(\R,i')| < d$ as needed. Let $V = \view_{\alpha}(\R,j)$, let $p$ (resp. $p'$) be the minimal (resp. maximal) element of $L(T(j))$ in the order $<^{\pi}_{\alpha}$, and let $j'$ denote the least common ancestor of $p$ and $p'$ in $T$, with $j' \leq_T j$. For each $x \in V$, choose an element $p_x \in L(T(x))$ such that $p <^{\pi}_{\alpha} p_x <^{\pi}_{\alpha} p'$; this is possible as $R(\pi,T,x)$ and $R(\pi,T,j)$ $\alpha$-view each other, and as we cannot have $I_{\alpha}(\pi,T,j) \subset I_{\alpha}(\pi,T,x)$ by definition of $j$. Let $Y = \{ p_x \mid x \in V \}$. As the nodes of $V$ form an antichain for the relation $<_T$, the elements $p_x$ are distinct, implying that $|Y| = |V|$. Consider the set $X = \{p,p'\} \cup Y$, and let $T' = T | X$. By definition of $j'$, it is still a node of $T'$. Furthermore, we have that each $p_x$ is not in $L(T(j))$ and thus not in $L(T'(j'))$, implying that $L(T'(j')) = \{p,p'\}$. We claim that $j'$ is the internal node of $T'$ with minimum index. By way of contradiction, suppose that $j'$ is preceded by another internal node $k$, such that $k \leq j' \leq j \leq i$. Consider $k' \in S$ such that $k \leq_T k'$. Then $L(T(k'))$ intersects $Y$ and thus $R(\pi,T,k')$ and $R(\pi,T,j)$ $\alpha$-view each other, implying that $k' \in V$. It follows that $L(T(k'))$ contains two elements $p_x,p_y$, a contradiction. We obtain that $j'$ is the internal node of $T'$ with minimum index, and thus $L(T'(j')) = \{p,p'\}$ is a $d$-close pair of $\pi|X$. As $\Int_{\alpha}(\pi|X,p,p') = Y$ and $|V| = |Y|$, we conclude that $|V| < d$.

$(i) \Rightarrow (ii)$: The fact that $T$ verifies property (ii) is a consequence of the following points.

Point 1: let $i$ be the internal node of $T$ with minimum index, then $L(T(i))$ is a $d$-close pair in $\pi$. Suppose that $L(T(i)) = \{x,y\}$. Since the leaves of $T$ precede the internal nodes, $S(T,i) - \{i\}$ contains exactly the leaves of $T$ distinct from $x,y$. It follows that for each $\alpha \in \{1,2\}$, $\view_{\alpha}(\R(\pi,T,i),i)$ corresponds to the elements of $\Int_{\alpha}(\pi,x,y)$, which has thus cardinality less than $d$.

Point 2: for every $X \subseteq S(\pi)$, $\D(\pi | X, T | X)$ is a $d$-wide decomposition of $\pi | X$. It is enough to show this for $X$ of the form $S(\pi) - \{j\}$ with $|X| \geq 2$. Suppose that $X$ has this form, let $\pi' = \pi | X$ and let $T' = T | X$. Let $u$ denote the parent of leaf $j$ in $T$, let $v$ denote the other child of $u$, and let $w$ denote the parent of $u$ in $T$ (possibly undefined if $u$ is the root of $T$). Note that $T'$ is obtained from $T$ by suppressing the nodes $j$ and $u$, and attaching $v$ as a child of $w$. Fix $i \in I(T')$, let the associated sets be $S = S(T,i)$ and $S' = S(T',i)$, and let the associated rectangle families be $\R = \R(\pi,T,i)$ and $\R' = \R(\pi',T',i)$. We need to show that $\view(\R',i) < d$. Fix $\alpha \in \{1,2\}$, let $V = \view_{\alpha}(\R,i)$ and $V' = \view_{\alpha}(\R',i)$. Observe that for $k \in I(T')$, if the intervals $I_{\alpha}(\pi',\R',k)$ and $I_{\alpha}(\pi',\R',i)$ intersect, then in $\pi$ the corresponding intervals $I_{\alpha}(\pi,R,k)$ and $I_{\alpha}(\pi,\R,i)$ also intersect. We consider three cases:
\begin{itemize}
\item Case 1: $j \in S$. In this case, it holds that $S' = S - \{j\}$, which implies that $V' \subseteq V$ and thus $|V'| \leq |V|$.
\item Case 2: $u \in S$. In this case, it holds that $S' = S - \{u\} + \{v\}$. Thus, we have either $V' = V$ (if $u \notin V$) or $V' \subseteq V - \{u\} + \{v\}$ (if $u \in V$), and thus $|V'| \leq |V|$.
\item Case 3: $u,j \notin S$. In this case, it holds that $S' = S$, which implies that $V' \subseteq V$ and thus $|V'| \leq |V|$.
\end{itemize}
In all cases, we obtain that $|V'| \leq |V| < d$, which concludes the proof.
\end{proof}

\fi


\section{Finding decompositions}\label{sec:grid-obstructions}

We present in this section a linear-time algorithm that either finds a large grid or gives a decomposition of bounded width:
\begin{theorem} \label{t:algo-decomp} There exists an algorithm that,
  given a reduced permutation $\pi$ of length $n$, runs in $O(n)$ time, and
  either finds an $r \times r$-grid of $\pi$, or
  returns the merge sequence of a $g(r)$-wide decomposition of $\pi$,
  where $g(r) = 2^{O(r \log r)}$.
\end{theorem}
On one hand, Theorem~\ref{t:algo-decomp} proves that grids are the
only obstructions for having a bounded-width decomposition. On the other hand, this decomposition algorithm together with the algorithm of Section~\ref{sec:solving-subpattern} working on bounded-width decompositions show that \textsc{Permutation Pattern} is linear-time solvable for fixed $\ell$.

The proof of Theorem~\ref{t:algo-decomp} relies on the following
statement, which is a variation of the main technical result of Marcus
and Tardos~\cite{DBLP:journals/jct/MarcusT04} in the proof of the
Stanley-Wilf conjecture.
\begin{theorem} \label{t:marcus-tardos} Let $f(r) = r^4 \binom{r^2}{r}$. For every $p,q,r \in \nat$ with $p+q > 2$, if $M$ is a point set included in $[p] \times [q]$ with $|M| > f(r) (p+q-2)$, then $M$ contains an $r \times r$-grid. Moreover, such a grid can be found in time $O(|M|)$.
\end{theorem}

\iffull
As the result in \cite{DBLP:journals/jct/MarcusT04}
is not stated algorithmically and it finds a permutation pattern rather than a grid, we reproduce
the proof in Appendix~\ref{sec:appendix} with appropriate modifications.
\fi
 The proof of Theorem
\ref{t:algo-decomp} below yields $g(r) = 4f(r)$; therefore, any improvement to Theorem
\ref{t:marcus-tardos} would immediately improve Theorem \ref{t:algo-decomp}.

\begin{proof}[Proof (of Theorem~\ref{t:algo-decomp})] Let $d = 4 f(r)$. The algorithm (see Algorithm~\ref{alg:build}) maintains an integer $k$, a rectangle family
  $\R$, a merge sequence $\Sigma$ and a gridding $G$.
 Given a column $x$ of
  $G$ (resp. a row $y$ of $G$), we denote by $d_1(x)$
  (resp. $d_2(y)$) the number of rectangles of $\R$ included in column
  $x$ (resp. row $y$).
\newcommand{\Rows}{\vname{Rows}}
\newcommand{\Cols}{\vname{Cols}}
\newcommand{\LargeCells}{\vname{LargeCells}}
\newcommand{\coord}{\vname{coord}}
\newcommand{\size}{\vname{size}}
\newcommand{\cells}{\vname{cells}}
\newcommand{\link}{\vname{link}}
\newcommand{\linkCol}{\vname{linkCol}}
\newcommand{\rects}{\vname{rects}}
\newcommand{\row}{\vname{row}}
\newcommand{\col}{\vname{col}}
\newcommand{\linkRow}{\vname{linkRow}}
\newcommand{\last}{\vname{last}}

 Initially: $k = n+1$; $\R = \pi$; $\Sigma$ is empty; the
  gridding $G$ consists of rows $r_1,\ldots,r_s$ and columns
  $c_1,\ldots,c_s$, such that each row $r_i$ and each column $c_i$ ($1
  \leq i < s$) contains exactly $d$ points of $\pi$. The algorithm ensures that the following invariant conditions hold at each step:
\begin{enumerate}[label=(C\arabic*),noitemsep]
\item each rectangle of $\R$ is included in a cell of $G$;\label{inv1}
\item  for any column $x$ of $G$, $d_1(x) \leq d$, and for any row $y$ of $G$, we have $d_2(y) \leq d$;\label{inv2}
\item for any two consecutive columns $x,x'$ of $G$, we have $d_1(x) + d_1(x') > d$;\label{inv3}
\item for any two consecutive rows $y,y'$ of $G$, we have $d_2(y) + d_2(y') > d$;\label{inv4}
\item $\R$ is $d$-wide.\label{inv5}
\end{enumerate}
Clearly, these conditions hold initially.

The algorithm performs the following main step repeatedly. As long as
$\R$ contains at least two rectangles, it does the following: (i) it
looks for a cell $(x,y)$ of $G$ which contains at least two rectangles
of $\R$; (ii) if there is no such cell, then it constructs a point set
$M$ corresponding to the nonempty cells and invokes the algorithm of
Theorem~\ref{t:marcus-tardos} to find an $r \times r$-grid; (iii)
otherwise, let $i,j$ be two rectangles of $\R$ inside $G(x,y)$. The
algorithm merges them in a new rectangle numbered by $k$, i.e. it
updates $\R \leftarrow \R[i,j \rightarrow k]$, and it appends the pair
$(i,j,k)$ to $\Sigma$. After this merge, the algorithm can update the
gridding $G$ as follows: (i) if there is a column $x'$ of $G$ consecutive
to $x$ such that $d_1(x) + d_1(x') \leq d$, then merge columns $x$ and
$x'$; (ii) if there is a row $y'$ of $G$ consecutive to $y$ such
that $d_2(y) + d_2(y') \leq d$, then merge rows $y$ and
$y'$. Finally, the algorithm increments $k$, and moves to the next
step of the loop.
\begin{algorithm}[t]
\caption{$\textsc{BuildDecomposition}(\pi)$}\label{alg:build}
\textbf{Input:}\\
\ $\pi$: a permutation of length $n$
 \begin{algorithmic}[1]
\STATE $\R:=$ the rectangle family representing $\pi$
\STATE $\Sigma:=()$
\STATE $k:=n+1$
\STATE initialize gridding $G$ such that every row and column (except the last ones) contains exactly $d$ points
\STATE while $|\R|>1$
\STATE \xs if there are two rectangles $R(i),R(j)$ in some cell $G(x,y)$\label{step:main}
\STATE \xs\xs $\R:=\R[i,j\rightarrow k]$
\STATE \xs\xs append $(i,j,k)$ to $\Sigma$
\STATE \xs\xs if $d_1(x)+d_1(x')\leq d$ for some $x'\in \{x-1,x+1\}$
\STATE \xs\xs\xs\ merge columns $x$ and $x'$ in $G$
\STATE \xs\xs if $d_2(y)+d_1(y')\leq d$ for some $y'\in \{y-1,y+1\}$
\STATE \xs\xs\xs\ merge rows $y$ and $y'$ in $G$
\STATE \xs else
\STATE \xs\xs construct the point set $M$ \hspace{1cm}/* We have $|M|>f(r)(p+q-2)$ */\label{step:no}
\STATE \xs\xs use the algorithm of Theorem~\ref{t:marcus-tardos} to find an $r\times r$ grid in $M$\label{step:marcus-tardos}
\STATE \xs\xs return the grid
\STATE \xs$k:=k+1$
\STATE return $\Sigma$ \label{step:return}
\end{algorithmic}
\end{algorithm}

\textbf{Correctness.} To prove the correctness of the algorithm, we first observe that the
invariant conditions \ref{inv1}--\ref{inv5} hold every time
Step~\ref{step:main} of Algorithm~\ref{alg:build} is reached.  Indeed, \ref{inv1} remains true,
since we are modifying $G$ by merging rows and columns; \ref{inv2}
holds, since we merge two rows or columns only if they together
contain at most $d$ rectangles; and \ref{inv3}--\ref{inv4} hold, since
we immediately merge any pair of rows or columns that would violate
it.  Invariant \ref{inv5} is a consequence of \ref{inv1} and
\ref{inv2}: a rectangle can only view other rectangles in the
same row or column.

Suppose that $G$ is a
$p \times q$-gridding when Step~\ref{step:no} is reached, and let us
construct the point set $M = \{ (x,y) \in [p] \times [q] \mid G(x,y) \text{ contains
a rectangle of } \R \}$. As the condition in Step~\ref{step:main} did not
hold, each point $(x,y) \in M$ corresponds to a single rectangle of $\R$.
It follows that $|M| > d \lfloor \frac{p}{2} \rfloor \geq d \frac{p-1}{2}$ by
Invariant~\ref{inv3}, and $|M| > d \lfloor \frac{q}{2} \rfloor \geq d \frac{q-1}{2}$
by Invariant~\ref{inv4}. Thus, $|M| > d \frac{p+q-2}{4} = f(r) (p+q-2)$: we
obtain by Theorem \ref{t:marcus-tardos} that $M$ contains an $r \times r$-grid,
which yields an $r \times r$-grid in $\pi$. Therefore, Step~\ref{step:marcus-tardos} indeed finds an $r\times r$-grid in $\pi$, which we return.

Finally, we observe that the sequence $\Sigma$ returned in Step
\ref{step:return} is the merge sequence of a $d$-wide decomposition.
Indeed, these merges produce a sequence of rectangle families, with
the last one containing only a single rectangle. By invariant \ref{inv5},
each rectangle family is $d$-wide.
\iffull

\textbf{Implementation: data structures.} We now explain how to implement the algorithm in the claimed $O(n)$ running time. We will use an appropriate data structure to represent the gridding $G$. To ensure that the data structure uses linear space, we only reserve memory for non-empty cells of $G$, i.e. cells containing a rectangle of $\R$. To allow fast detection of a \emph{large} cell (i.e. containing at least two rectangles), the data structure also contains a list of the large cells. The data structure $\Gamma$ representing $G$ consists of:
\begin{itemize}
\item a linked list $\Rows$ storing pointers to the rows of $G$, sorted from bottom to top;
\item a linked list $\Cols$ storing pointers to the columns of $G$, sorted from left to right;
\item a linked list $\LargeCells$ storing pointers to the cells of $G$ which contain at least two rectangles;
\item for each row $p \in \Rows$, an integer $\coord(p)$ inducing an increasing numbering of the rows, an integer $\size(p)$ counting the number of rectangles inside row $p$, a linked list $\cells(p)$ storing pointers to the non-empty cells of row $p$, sorted from left to right, and a pointer $\link(p)$ to the element of list $\Rows$ corresponding to $p$;
\item for each column $p \in \Cols$, an integer $\coord(p)$ inducing an increasing numbering of the columns, an integer $\size(p)$ counting the number of rectangles inside column $c$, a linked list $\cells(p)$ storing pointers to the non-empty cells of column $p$, sorted from bottom to top, and a pointer $\link(p)$ to the element of list $\Cols$ corresponding to $p$;
\item for each cell $c$, pointers $\row(c)$ and $\col(c)$ to the row and column of $c$, a linked list $\rects(c)$ storing the indices of the rectangles inside $c$, a variable $\size(c)$ equal to the length of $\rects(c)$, a variable $\linkRow(c)$ pointing to the entry of $\cells(\row(c))$ representing $c$, a variable $\linkCol(c)$ pointing to the entry of $\cells(\col(c))$ representing $c$, and a variable $\link(c)$ pointing to the element of list $\LargeCells$ containing $c$ (or $\perp$ if $\LargeCells$ does not contain $c$).
\end{itemize} 
The data structure supports the following operations: (i) $\textsc{MergeRows}(\Gamma,p,p')$ takes two consecutive rows $p$ and $p'$, and merges them; (ii) $\textsc{MergeCols}(\Gamma,p,p')$ takes two consecutive columns $p$ and $p'$ and merges them. We briefly describe how to implement \textsc{MergeRows} in time $O(\size(p)+\size(p'))$. First, it allocates memory for a new row $p''$, and updates the list $\Rows$ to replace $p$ and $p'$ by $p''$. Second, it sets $\coord(p'')$ to $\coord(p)$, $\size(p'')$ to $\size(p) + \size(p')$, and $\link(p'')$ to the element of $\Rows$ corresponding to $p''$. Finally, it constructs the list $\cells(p'')$ by a simultaneous traversal of the lists $\cells(p)$ and $\cells(p')$. The traversal maintains a cell $c \in \cells(p)$ and a cell $c' \in \cells(p')$, starting at the first entry of each list. When reading cell $c \in \cells(p)$ and $c' \in \cells(p')$, the algorithm allocates a new cell $c''$ with $\row(c'') = p''$, appends $c''$ to $\cells(p'')$, sets $\linkRow(c'')$ accordingly, then proceeds as follows.
\begin{itemize}
\item Case 1: if the column of $c$ is to the left of the column of $c'$ (i.e. $\coord(\col(c)) < \coord(\col(c'))$). Then set $\col(c'')$ to $\col(c)$, $\rects(c'')$ to $\rects(c)$, $\size(c'')$ to $\size(c)$; replace $c$ by $c''$ in $\cells(\col(c))$ using the pointers $\linkCol$; update $\LargeCells$ by removing $c$ and adding $c''$ if $\size(c) > 1$; finally, update $c$ to the next cell of $\cells(p)$.
\item Case 2: if the column of $c$ is to the right of the column of $c'$ (i.e. $\coord(\col(c)) > \coord(\col(c'))$). This case is handled symmetrically to Case 1.
\item Case 3: if $c$ and $c'$ are in the same column (i.e. $\col(c) = \col(c')$). Then set $\col(c'')$ to $\col(c)$, $\rects(c'')$ to $\rects(c) \cup \rects(c')$, $\size(c'')$ to $\size(c) + \size(c')$; replace $c$ and $c'$ by $c''$ in $\cells(\col(c))$ using the pointers $\linkCol$; update $\LargeCells$ by removing $c$ and $c'$ if needed, and adding $c''$; finally, update $c$ to the next cell of $\cells(p)$, and $c'$ to the next cell of $\cells(p')$.
\end{itemize} 
\textbf{Implementation: details of the steps.}
 Given $\pi$, we first construct the initial data structure $\Gamma$ in time $O(n)$. For simplicity, we only describe how to construct the rows and the cells. We first allocate rows $r_1,\ldots,r_s$ and columns $c_1,\ldots,c_s$ with $s = \lceil \frac{n}{d} \rceil$, and we initialize $\coord(r_i)$ and $\coord(c_i)$ to $i$. We will construct the lists $\cells(r_y)$ by scanning the points of $\pi$ by increasing $x$-coordinate. To each row $r_y$, we associate a variable $\last[y]$ (initialized to $\perp$) which points to the last nonempty cell in row $r_y$ among the points already examined. When examining a new point $p$, we compute the index $x$ of its column and the index $y$ of its row. If $\last[y] \neq \perp$ and $x$ is the index of $\col(\last[y])$, then we add $p$ to $\rects(\last[y])$, and we update $\size(\last[y])$ and $\LargeCells$ if needed. Otherwise, we allocate a new cell $c$, we set $\row(c)$ to $r_y$, $\col(c)$ to $c_x$, $\size(c)$ to 1 and $\rects(c)$ to $\{p\}$. Then, we append $c$ to $\cells(r_y)$, we set $\linkRow(c)$ accordingly, and we update $\last[y]$ to $c$.

Now, each step of the main loop is accomplished as follows. If $\LargeCells$ is empty, then we compute the point set $M$ and invoke the algorithm of Theorem~\ref{t:marcus-tardos} to obtain an $r\times r$ grid. The construction of $M$ is accomplished by first renumbering the rows and columns with consecutive integers, and then by enumerating the cells and collecting their coordinates. If $\LargeCells$ is not empty, let $c$ be the first element of $\LargeCells$, let $p = \col(c)$ and $q = \row(c)$, and let $i,j$ be the first two elements of $\rects(c)$; replace $i,j$ by $k$ in $\rects(c)$, and decrement $\size(c)$, $\size(p)$ and $\size(q)$; if $\size(c)$ becomes equal to 1 then remove $c$ from $\LargeCells$; if there exists a column $p'$ consecutive to $p$ such that $\size(p') + \size(p) \leq d$, call $\textsc{MergeCols}(\Gamma,p,p')$; if there exists a row $q'$ consecutive to $q$ such that $\size(q') + \size(q) \leq d$, call $\textsc{MergeRows}(\Gamma,q,q')$. Observe that each step of the main loop takes $O(1)$ time, excluding the calls to \textsc{MergeXXX}. Next, observe that the number of these calls is equal to the initial number of rows and columns of $\Gamma$, which is $O(\frac{n}{d})$; as each such call takes $O(d)$ time, it follows that they take $O(n)$ time overall. Summing up the time taken by the initialization, by the main loop and by the calls to \textsc{MergeXXX}, we obtain that the total running time is $O(n)$. 
\fi
\end{proof}
We close this section by stating a corollary of Theorem \ref{t:algo-decomp}. Given a permutation $\pi$, we can define three values that measure the ``complexity'' of $\pi$. The first measure is the largest integer $r$ such that $\pi$ contains an $r \times r$-grid; we denote this measure as $g(\pi)$. The second measure is the smallest integer $d$ such that every subpattern of $\pi$ has a $d$-close pair; we denote this measure as $d(\pi)$. The third measure is the width of $\pi$ defined earlier, denoted by $w(\pi)$. We observe that these three measures are equivalent, in the following sense: we say that two functions $m,m'$ mapping permutations to integers are \emph{equivalent} if there exist increasing functions $f,g : \nat \rightarrow \nat$ such that $m(\pi) \leq f(m'(\pi))$ and $m'(\pi) \leq g(m(\pi))$ hold for any permutation $\pi$.

\begin{corollary} The measures $g, d$ and $w$ are equivalent.
\end{corollary}
\iffull
\begin{proof} The equivalence between $g$ and $w$ follows from Proposition \ref{p:gridwidth} and Theorem \ref{t:algo-decomp}. We now argue that $d$ is equivalent to the other two, by showing that $d(\pi) \leq w(\pi)$ and $g(\pi) < 5 d(\pi)$. On one hand, $d(\pi) \leq w(\pi)$ follows from Proposition \ref{p:alt-charac}. On the other hand, $g(\pi) < 5 d(\pi)$ follows by showing that any $5r \times 5r$-grid permutation $\pi$ has a subpattern containing no $r$-close pair. Suppose that $\pi$ is such a permutation, let $G$ be the corresponding $5r \times 5r$-gridding, and let $p(x,y)$ denote the point of $\pi$ in column $x$, row $y$ of $G$. We define the subset $S' \subseteq S(\pi)$ containing the points $p(x,y)$ such that $y \equiv 2x\pmod{5}$, and we let $\pi' = \pi | S'$. Observe that $S'$ contains exactly $r$ points in each row and column of $G$. Furthermore, we cannot have two distinct points $p(x,y), p(x',y') \in S'$ with $|x'-x| \le 1$ and $|y'-y| \le 1$. It follows that for any two elements $p,p' \in S'$, either $|\Int_1(\pi',p,p')| \ge r$ (if $p,p'$ belong to non-consecutive columns) or $|\Int_2(\pi',p,p')| \ge r$ (if $p,p'$ belong to non-consecutive rows). We conclude that $\pi'$ is a subpattern of $\pi$ with no $r$-close pair.
\end{proof}
\fi


\newcommand{\Rng}{\textup{Rng}}
\newcommand{\VEndpoints}{\vname{VEndpoints}}
\newcommand{\HEndpoints}{\vname{HEndpoints}}
\newcommand{\rect}{\vname{rect}}
\newcommand{\Rects}{\vname{Rects}}
\newcommand{\Intt}{\vname{Int}}
\newcommand{\coord}{\vname{coord}}

\section{Solving the Permutation Pattern problem}\label{sec:solving-subpattern}
This section is devoted to showing that the \textsc{Permutation Pattern} problem can be solved in linear time if a decomposition of bounded width is given in the input:
\begin{theorem}\label{th:dp} The \textsc{Permutation Pattern} problem can be solved
  in time $(d\ell)^{O(\ell)}\cdot n$, where $\ell=|\sigma|$ and
  $n=|\pi|$, if the merge sequence of a $d$-wide decomposition of
  $\pi$ is given in the input.
\end{theorem}
To prove Theorem~\ref{th:main}, we run first the algorithm of
Theorem~\ref{t:algo-decomp} on permutation $\pi$ with $r = \ell$,
which takes $O(n)$ time.  If the algorithm concludes that $\pi$ has an
$\ell \times \ell$-grid, we conclude that $\sigma$ is a subpattern of
$\pi$ and we answer ``yes''. Otherwise, we obtain the merge sequence
of a $g(\ell)$-wide decomposition of $\pi$, where
$g(\ell)=2^{O(\ell \log \ell)}$. Using Theorem~\ref{th:dp}, we can
then decide if $\sigma$ is a subpattern of $\pi$ in time
$(g(\ell)\ell)^{O(\ell)}\cdot n=2^{O(\ell^2 \log\ell)}\cdot n$.

  For the proof of Theorem~\ref{th:dp}, let $\D=(\R_0,\dots,
  \R_{n-1})$ be the decomposition of $\pi$ given in the input, with
  $\R_i=(S_i,R_i)$. Recall that each rectangle in $\R_i$ was created
  by a sequences of merges (possibly 0) from the rectangle family $\R_0$
  representing the permutation $\pi$. We denote by $L(j)$ the set of
  points (more precisely, indices) taking part in the merges creating rectangle
  indexed by $j$. For example, in Figure~\ref{fig:merge}, we have
  $L(13)=\{3,4,5,8\}$. Note that, even if a point $p$ is covered by
  rectangle $j$, it is not necessarily in $L(j)$: for example, in
  Figure~\ref{fig:merge}, point 6 is covered by rectangle 13, but 6 is
  not in $L(13)$, as it did not take part in any of the 3 merges
  creating 13 (point 6 appears only in $L(6)$, $L(11)$, $L(14)$, and
  $L(15)$).

\begin{figure}
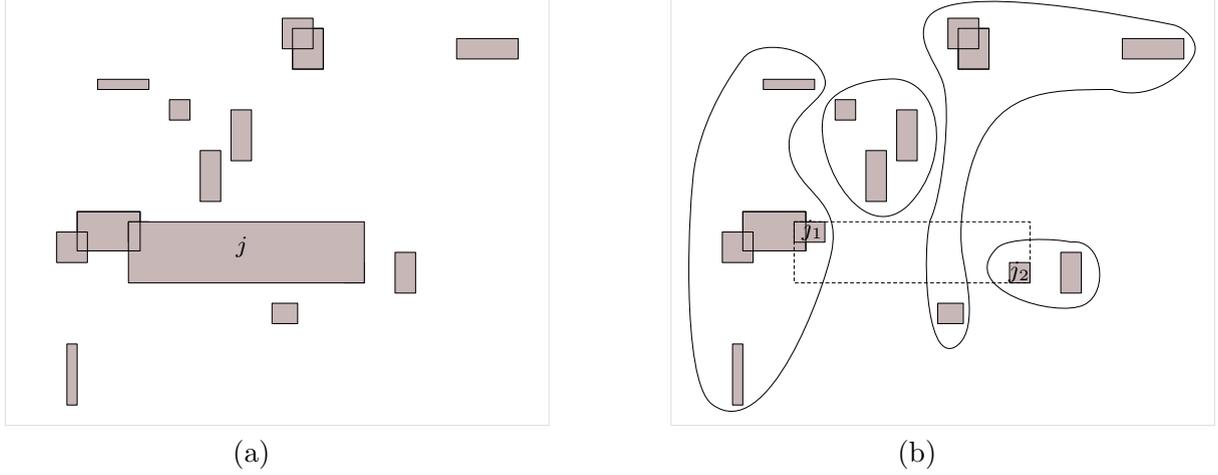

\begin{center}
{\footnotesize \svg{0.45\linewidth}{conn1}
\hfill
\svg{0.45\linewidth}{conn2}}
\caption{Step $i+1$ of the decomposition merges $j_1$ and $j_2$ to $j$. (a) A connected set $K$ of $G_{i+1}$ containing $j$. (b) Replacing $j$ with $j_1$ and $j_2$ gives a set $K^{\Pi}$ inducing 4 connected components in $G_i$.}\label{fig:conn}
\end{center}
\end{figure}

 For $0 \le i \le n-1$, we define the
{\em visibility graph} $G_i$ at step $i$ of the decomposition the
following way: the vertex set of $G_i$ is $S_i$ and $x,y\in S_i$ are
adjacent if and only if the rectangles $R_i(x)$ and $R_i(y)$
$\alpha$-view each other for some $\alpha\in \{1,2\}$. As $\R_i$ is a
$d$-wide rectangle family, it follows that $G_i$ has maximum degree at
most $2d$. Figure~\ref{fig:conn}(a) shows a connected set of
rectangles in the visibility graph (by ``connected set'', we mean that they induce a connected subgraph of the visibility graph).

 We solve the \textsc{Permutation Pattern} problem using dynamic
 programming.  For each step $i$, we define a set of
 subproblems. Informally, a subproblem asks for a subset of $\sigma$
 to be embedded into elements of $\pi$ that appear in a set $K$ of rectangles
 inducing a connected subgraph of the visibility graph $G_i$, with the elements of $\sigma$ distributed among the rectangles of $K$ in a specified way.

 For the formal definition of the subproblems, we need the following
 definition first.  Given two sets $X,Y$, a \emph{distribution of $X$
   into $Y$} is a function $F : Y \rightarrow 2^X$ such that for $i,j
 \in Y$ distinct, $F(i) \cap F(j) = \emptyset$; the \emph{range} of
 $F$ is $\Rng(F) = \cup_{i \in Y} F(i)$.  An \emph{admissible
   subproblem} is a triple $t = (i,K,F)$, where
\begin{itemize}[noitemsep]
\item $0\le i \le n-1$,
\item  $K$ is a connected subset of $G_i$, and 
\item $F$ is a distribution of $S(\sigma)$ into $K$ such that $F(i) \neq \emptyset$ for each $i \in K$.
\end{itemize}
Note that the last condition implies that $K$ can have at most $\ell$ vertices. We define the \emph{range} of $t$ as $\Rng(F)$.
The number of possible distributions is $(|K|+1)^{|S(\sigma)|}=\ell^{O(\ell)}$.
The following simple fact bounds the number of possible connected sets, and it follows that the number of subproblems is $(d\ell)^{O(\ell)}\cdot n$ for a given $i$.
\begin{proposition}\label{lem:connectedbound}
  If $G$ is a graph with maximum degree $\Delta$ and $v$ is a vertex
  of $G$, then the number of sets $K$ of size at most $\ell$ such that $v\in K$
  and $G[K]$ is connected is $\Delta^{O(\ell)}$. Moreover,\iffull\ all\fi\ these sets can be
  enumerated in time $\Delta^{O(\ell)}$.
\end{proposition}
\iffull
\begin{proof}
  The vertices of $G[K]$ can be visited by a walk of length at most
  $2\ell-1$ starting at $v$. In each step of the walk, we move to one
  of the at most $\Delta$ neighbors. Thus there are at most $\Delta^{2\ell-1}$ such
  walks, which is an upper bound on the number of sets $K$. Enumerating all these walks gives $\Delta^{O(\ell)}$ sets that are
  not necessarily distinct. However, we may sort these sets in time
  $\Delta^{O(\ell)}$ and remove the duplicates.
\end{proof}
\fi

 We say that $t$ is \emph{satisfiable} iff there exists a mapping $\phi : \Rng(F) \rightarrow S(\pi)$ such that
\begin{enumerate}[noitemsep]
\item[(i)] for each $p \in \Rng(F)$, if $p \in F(j)$, then $\phi(p) \in L(j)$, and
\item[(ii)] $\phi$ is an embedding of $\sigma | \Rng(F)$ into $\pi$.
\end{enumerate} 
In this case, we say that $\phi$ is a {\em solution} of $t$.  Recall
that $\R_{n-1}$ contains only a single rectangle $j$ and
$L(j)=S(\pi)$. Therefore, there is an embedding from $\sigma$ to $\pi$
if and only if the subproblem $(n-1,\{j\},F)$ is satisfiable, where $F$ is
the distribution of $S(\sigma)$ into $\{j\}$ such that $F(j)=S(\sigma)$.

Lemma \ref{l:recurrence} below gives a recurrence relation that allows
us to decide if a subproblem $t = (i+1,K,F)$ is satisfiable, assuming
that we have computed the satisfiable subproblems at step $i$. Our
goal is to show that a solution $\phi$ for $t$ can be constructed by
putting together solutions for particular subproblems at step $i$.

Suppose that $t = (i+1,K,F)$ is an admissible subproblem, and suppose
that step $i+1$ of the decomposition merges $j_1,j_2$ into $j$. Clearly, this means that $L(j)$ is the disjoint union of
$L(j_1)$ and $L(j_2)$. Any solution $\phi$ of $t$ maps
$F(j)$ to $L(j)=L(j_1)\cup L(j_2)$, hence it defines a bipartition of the
elements of $L(j)$. As a first step of solving $t$, we guess this
bipartition, that is, which elements of $F(j)$ are mapped to $L(j_1)$
and to $L(j_2)$ (there are $2^{|L(j)|}\le 2^\ell$ such bipartitions).
Let $X = F(j)$, and fix a bipartition $\Pi = (X_1,X_2)$ of
$X$. Mapping $\phi$ maps $\Rng(F)$ to the rectangles
$K-\{j\}+\{j_1,j_2\}$ of $G_i$. However, there is a technical detail
here: if $X_1$ or $X_2$ is empty, then $\phi$ does not map any element of $\Rng(F)$ 
to $L(j_1)$ or $L(j_2)$, respectively. Therefore, we define the set $K^{\Pi}$ as
follows:
\[
K^{\Pi}=\begin{cases}
K - \{j\} +
\{j_1,j_2\} & \text{if $X_1,X_2 \neq \emptyset$,}\\
K-\{j\}+\{j_1\} & \text{if $X_1\neq \emptyset$, $X_2=\emptyset$,}\\
K-\{j\}+\{j_2\} & \text{if $X_1=\emptyset$, $X_2\neq\emptyset$.}
\end{cases}
\]
 We define the distribution $F^{\Pi}:  K^{\Pi}
\to 2^{S(\sigma)}$ that describes how mapping $\phi$ maps the elements of $K$ to the rectangles in $K^{\Pi}$:
\[
F^{\Pi}(k) = 
\begin{cases}
F(k) & \text{if $k\not\in \{j_1,j_2\}$},\\
X_1 & \text{if $k=j_1$,}\\
X_2 & \text{if $k=j_2$.}\\
\end{cases}
\]
Assuming that we have already computed the satisfiable subproblems at
step $i$, we would like to use this information to decide whether
there is a solution $\phi$ satisfying $t=(i+1,K,F)$ that corresponds
to the bipartition $\Pi$. Let us observe that if $(i,K^{\Pi},F^{\Pi})$
happens to be an admissible and satisfiable subproblem, then it
immediately implies the existence of such a solution $\phi$. However,
in general, $K^{\Pi}$ is not necessarily connected. In that case, we
would like to put together the solution $\phi$ from the solutions of
the subproblems corresponding to the connected components of
$G_i[K^{\Pi}]$.  Formally, if $C$ is a connected component of
$G_i[K^{\Pi}]$, we define the subproblem $t^{\Pi,C} = (i,C,F^{\Pi} |
C)$. If $C_1$, $\dots$, $C_m$ are the connected components of
$G_i[K^{\Pi}]$, then we let $T^{\Pi}=\{t^{\Pi,C_1},\dots,
t^{\Pi,C_m}\}$ denote the set of subproblems corresponding to these
components; observe that they are admissible subproblems. Note that
$G_i[K^{\Pi}]$ can have more than two connected components: it is not
true that every connected component contains either $j_1$ or $j_2$. As
Figure~\ref{fig:conn}(b) shows, rectangle $j$ can view rectangles that
neither $j_1$ nor $j_2$ view, thus there can be connected components
not containing either $j_1$ or $j_2$.

We would like to combine solutions of the subproblems in $T^{\Pi}$ to
obtain a solution for subproblem $t$. However, the subproblems have to
satisfy a certain condition in order for this to be possible.  Consider
two admissible subproblems $t_1 = (i,K_1,F_1)$ and $t_2 =
(i,K_2,F_2)$. We say that $t_1$ and $t_2$ are \emph{independent} if
$\Rng(F_1), \Rng(F_2)$ are disjoint, $K_1$ and $K_2$ are disjoint, and
$G_i$ has no edge between $K_1$ and $K_2$. Observe that the
subproblems in $T^{\Pi}$ are pairwise independent. Suppose that $t_1$
and $t_2$ are independent with solutions $\phi_1:\Rng(F_1)\to S(\pi)$
and $\phi_2:\Rng(F_2)\to S(\pi)$,
respectively.  Then the mapping $\phi:\Rng(F_1)\cup \Rng(F_2)\to
S(\pi)$ defined the obvious way from $\phi_1$ and $\phi_2$ is not
necessarily a correct embedding of $\sigma|\Rng(F_1)\cup \Rng(F_2)$. The
problem is that if $r_1\in K_1$ and $r_2\in K_2$, then for some
$s_1\in F_1(r_1)$ and $s_2\in F_2(r_2)$, the relative position of the
points $\phi(s_1)$ and $\phi(s_2)$ is not necessarily the same as the
relative position of $s_1$ and $s_2$. However, the crucial observation
here is that the rectangles $r_1$ and $r_2$ do not view each other,
hence the relative position of $\phi(s_1)$ and $\phi(s_2)$ depend only
on the relative position of $r_1$ and $r_2$, and {\em not} on the
actual selection of points in $r_1$ and $r_2$. Therefore, the sets
$K_1$ and $K_2$ and the distributions $F_1$ and $F_2$ already
determine if the two solutions can be combined. Formally, two
independent subproblems $t_1 = (i,K_1,F_1)$ and $t_2 = (i,K_2,F_2)$,
are said to be \emph{compatible} if for each $r_1 \in K_1, r_2 \in K_2$ and $\alpha \in \{1,2\}$:
\begin{itemize}[noitemsep]
\item if $I_{\alpha}(R_i(r_1)) < I_{\alpha}(R_i(r_2))$ then for each $p_1 \in F_1(r_1), p_2 \in F_2(r_2)$ it holds that $p_1
<^{\sigma}_{\alpha} p_2$, and
\item if $I_{\alpha}(R_i(r_2)) < I_{\alpha}(R_i(r_2))$ then for each $p_1 \in F_1(r_1), p_2 \in F_2(r_2)$ it holds that $p_2 <^{\sigma}_{\alpha} p_1$.
\end{itemize}

Note that by the independence assumption for $t_1$ and $t_2$, one of the above two conditions must hold.
We are now able to state the recurrence relation as follows:

\begin{lemma} \label{l:recurrence} Suppose that $t = (i+1,K,F)$ is an admissible subproblem. Then $t$ is satisfiable \iffull if and only if\fi
\begin{itemize}[noitemsep]
\item $j \notin K$ and $t' = (i,K,F)$ is satisfiable, or
\item $j \in K$ and there exists a bipartition $\Pi$ of $X$ such that
  the subproblems of $T^{\Pi}$ are satisfiable and pairwise
  compatible.
\end{itemize}
\end{lemma}
\iffull
\begin{proof} The case when $j\notin K$ is clear, by observing that if $j \notin K$ then $t'$ is also an admissible subproblem. Let us consider the case when $j\in K$.

Suppose that $t$ is satisfiable through a mapping $\phi : Z \rightarrow S(\pi)$ with $Z$ the range of $t$. Let $\Pi = (X_1,X_2)$ with $X_r = \phi^{-1}(L(j_r))$, we show that the subproblems of $T^{\Pi}$ satisfy the requirement. Suppose that the connected components of $G_i[S^{\Pi}]$ are $C_1,\ldots,C_m$, and let $t_r = t^{\Pi,C_r}$ for $r \in [m]$. As the $C_r$'s are the connected components of $G_i[S^{\Pi}]$, each $t_r$ is an admissible subproblem, and they are pairwise independent. Let $Z_r$ denote the range of $t_r$, then $Z_1,\ldots,Z_m$ form a partition of $Z$. We first show that each subproblem $t_r$ is satisfiable through $\phi | Z_r$. Point (ii) of the definition is verified as it holds for $\phi$, and for point (i) observe that if $p \in F^{\Pi}(k)$ and $k \in C_r$ then: either $k \neq j_1,j_2$, hence $p \in F(k)$ and thus $\phi(p) \in L(k)$ (by definition of $\phi$), or $k = j_s$ in which case $p \in X_s$ and thus $\phi(p) \in L(j_s)$ (by definition of $X_s$). We now show that two triples $t_r = (i,K_1,F_1), t_s = (i,K_2,F_2)$ are compatible. Indeed, suppose that $k \in K_1$ and $k' \in K_2$ are such that $I_{\alpha}(R_i(k)) < I_{\alpha}(R_i(k'))$, then for $p \in F_1(k)$ and $p' \in F_2(k')$ we have $\phi(p) \in L(k)$, $\phi(p') \in L(k')$ (by definition of $\phi$), and thus $\phi(p) <^{\pi}_{\alpha} \phi(p')$, which implies that $p <^{\sigma}_{\alpha} p'$ as $\phi$ is an embedding.

Conversely, suppose that there exists $\Pi = (X_1,X_2)$ bipartition of $X$ such that the subproblems of $T^{\Pi}$ are satisfiable and pairwise compatible. Suppose that the connected components of $G_i[S^{\Pi}]$ are $C_1,\ldots,C_m$, and let $t_r = t^{\Pi,C_r}$ for $r \in [m]$. Let $Z_r$ denote the range of $t_r$, and let $Z$ denote the range of $t$, then $Z_1,\ldots,Z_m$ form a partition of $Z$. Suppose that $t_r$ is satisfiable through a mapping $\phi_r : Z_r \rightarrow S(\pi)$. We can then define $\phi : Z \rightarrow S(\pi)$ which coincides with $\phi_r$ on $Z_r$. We show that $t$ is satisfiable through $\phi$. We first show Point (i) of the definition. Suppose that $p \in F(k)$. If $k \neq j$, we have that $p \in (F^{\Pi} | C_r)(k)$ for some $r$, which implies that $\phi(p) = \phi_r(p) \in L(k)$ (by definition of $\phi_r$). If $k = j$, then we have $p \in X_s$ for some $s \in \{1,2\}$, and thus $p \in (F^{\Pi} | C_r)(j_s)$ for the component $C_r$ containing $j_s$, which implies that $\phi(p) = \phi_r(p) \in L(j_s)$ (by definition of $\phi_r$) and thus in $L(j)$. We now show Point (ii) of the definition. Let $p,p'$ be two elements of $Z$ and $\alpha \in \{1,2\}$, and suppose that $p <^{\sigma}_{\alpha} p'$, we need to show that $\phi(p) <^{\pi}_{\alpha} \phi(p')$. If $p,p'$ belong to a same set $Z_r$, then $\phi(p) = \phi_r(p) <^{\pi}_{\alpha} \phi_r(p') = \phi(p')$ (by definition of $\phi_r$). If $p \in Z_r, p' \in Z_s$ with $r \neq s$, suppose that $p \in (F^{\Pi} | C_r)(k)$ and $p' \in (F^{\Pi} | C_s)(k')$. As $t_r$ and $t_s$ are compatible and as $p <^{\sigma}_{\alpha} p'$, we have $I_{\alpha}(R_i(k)) < I_{\alpha}(R_i(k'))$. As $\phi(p) = \phi_r(p) \in L(k)$ and $\phi(p') = \phi_s(p') \in L(k')$, we conclude that $\phi(p) <^{\pi}_{\alpha} \phi(p')$. 
\end{proof}
\fi
\begin{algorithm}[t]
\caption{$\textsc{FindPattern}(\pi,\sigma,\D)$}\label{alg:find}
\textbf{Input:}\\
\ $\pi$: a permutation of length $n$\\
\ $\sigma$: a permutation of length $\ell$\\
\ $\D$: a decomposition of $\pi$
 \begin{algorithmic}[1]
\STATE for $i:=0 \text{ to } n-2$
\STATE \xs suppose that step $i+1$ merges $j_1$, $j_2$ to obtain $j$.
\STATE \xs update $G_i$ to obtain $G_{i+1}$.
\STATE \xs for every subproblem $t=(i+1,K,F)$
\STATE \xs\xs if $j \notin K$
\STATE \xs\xs\xs add 
$(i+1,K,F)$ to the list of satisfiable subproblems if $(i,K,F)$ is satisfiable.
\STATE \xs\xs else
\STATE \xs\xs\xs for every bipartition $\Pi$ of $F(j)$
\STATE \xs\xs\xs\xs compute $V^{\Pi}$
\STATE \xs\xs\xs\xs compute the connected components $C_1$, $\dots$, $C_{q}$
\STATE \xs\xs\xs\xs compute the set $T^{\Pi}$ of subproblems.
\STATE \xs\xs\xs\xs if the subproblems in $T^{\Pi}$ are pairwise compatible
\STATE \xs\xs\xs\xs\xs add $(i+1,K,F)$ to the list of satisfiable subproblems.
\STATE let $j$ be the unique rectangle in $\R_{n-1}$
\STATE let $F$ be the distribution of $S(\sigma)$ into $\{j\}$ with $F(j)=S(\sigma)$.
\STATE if $(n-1,j,F)$ is satisfiable
\STATE \xs return ``yes''
\STATE else
\STATE \xs return ``no''
\end{algorithmic}
\end{algorithm}

Lemma~\ref{l:recurrence} allows us to determine if $t=(i+1,K,F)$ is
satisfiable, assuming that we have solved all subproblems at step
$i$ (see Algorithm~\ref{alg:find}). Let us briefly sketch how to implement this algorithm in time
$(d \ell)^{O(\ell)}\cdot n$.

We maintain a representation of the graph $G_i$ and for every $v\in
S_i$, a linked list of every subset $K$ of size at most $\ell$
containing $v$ such that $G_i[K]$ is connected; note that each such
list is of size $d^{O(\ell)}$ by Lemma~\ref{lem:connectedbound}. For each such set $K$, there is a linked
list of all the satisfiable subproblems $(i,K,F)$ for every
distribution $F$; the length of this list is $\ell^{O(\ell)}$.  In
order to get $G_{i+1}$ from $G_{i}$ efficiently, we maintain a sorted
linked list of the horizontal endpoints of all the rectangles
appearing in $\R_i$ (containing two entries for each rectangle) and a
similar list for the vertical endpoints. When $j_1$ and $j_2$ are
merged to obtain $j$, we can use these lists of endpoints to efficiently find all
those rectangles that $j$ views, but neither $j_1$ not $j_2$
views. Therefore, we can efficiently update $G_i$ to obtain
$G_{i+1}$. Then for every vertex of $G_{i+1}$ at distance at most
$\ell$, we have to recompute the list of connected subsets; there are
$d^{O(\ell)}$ such vertices and enumeration of the subsets takes
$d^{O(\ell)}$ time at each vertex.

To solve the subproblems at step $i+1$, we have to update only those
subproblems $t=(i+1,K,F)$ for which $j\in K$ holds, where $j$ is the
new rectangle created at this step. For each such subproblem, we
enumerate every bipartition $\Pi$ of $L(j)$ and compute the set
$K^{\Pi}$ and distribution $F^{\Pi}$. We compute the connected
components $C_1$, $\dots$, $C_m$ of $G_i[K^{\Pi}]$ and the set of
subproblems $T^{\Pi}$. By Lemma~\ref{l:recurrence}, we need to check
if these problems are pairwise compatible, and if so, we have to add
this satisfiable subproblem to the list of every $v\in K$. Note that
all these steps involve only vertices at distance at most $\ell$ from
$j$, $j_1$, or $j_2$, whose number is $d^{O(\ell)}$, and each operation
  can be performed in time $d^{O(\ell)}$. As there are
  $(d\ell)^{O(\ell)}$ subproblems $(i+1,K,F)$ with $j\in K$, all the
  updates for step $i+1$ can be done in time $(d\ell)^{O(\ell)}$.

\iffull
\subsection{Implementation}
\label{sec:dpimplementation}

In this section, we describe an improved version of the previous algorithm for which we obtain a precise running time of $O((16 \ell d)^{\ell} \ell^2 n)$. We make several optimizations: (i) we use a more efficient way of enumerating the connected subsets of $G_i$; (ii) we arrange the subproblems in a table such that any subproblem can be accessed in constant time; (iii) we ensure that when considering a step $i$ which produces a rectangle $j$, we only have to compute the table entries for $j$, without having to update the other table entries.

Let us first describe optimization (i). The proof of Proposition \ref{lem:connectedbound} gives a crude upper bound of $\Delta^{2\ell-1}$ which can be improved as follows. We say that a function $N_i : S_i \times [2d] \rightarrow S_i \cup \{\perp\}$ is an \emph{adjacency function} for graph $G_i$ iff for each $k \in S_i$, it holds that $\{N_i(k,p) : 1 \leq p \leq 2d\} - \{\perp\}$ equals the neighborhood of $k$ in $G_i$. We let $\T$ denote the set of rooted trees with at most $\ell$ nodes, whose vertices are numbered by a prefix ordering, and whose edges are labeled by integers in $[2d]$; observe that $\T$ has size $O((8d)^\ell)$ and can be constructed in linear time. Consider an adjacency function $N_i$ for $G_i$. Given a vertex $k \in S_i$ and a tree $T \in \T$, we define a subgraph $G' \subseteq G_i$ as follows: start with $G'$ containing the single vertex $k$, $U = \{1\}$, and assign $k$ to the vertex $1$ of $T$; while $T$ contains an arc $(x,y)$ with $x \in U, y \notin U$ having label $p$, let $j$ be the vertex associated to $x$, let $j' = N_i(j,p)$, fail if $j' = \perp$, otherwise assign $j'$ to the vertex $y$, add the edge $j j'$ to $G'$ and add $y$ to $U$. If the algorithm fails at some point, we set $\Phi(N_i,k,T) = \perp$, otherwise $\Phi(N_i,k,T)$ is the resulting graph $G'$. Observe that for every connected subset $K$ of $G_i$ with at most $\ell$ vertices and containing $k$, there is some $T \in \T$ such that $\Phi(N_i,k,T)$ is a spanning tree of $G_i[K]$. This immediately gives an upper bound of $O((8d)^{\ell})$ on the number of such sets $K$. The algorithm will maintain at step $i$ an adjacency function $N_i$ from which these subsets can be determined.

We now describe optimization (ii). The previous algorithm maintained for each vertex $v \in S_i$, a linked list of the admissible subproblems $(i,K,F)$ with $v \in K$. This required traversing the list in time $(d \ell)^{O(\ell)}$ to access a given subproblem. Instead, we use a single table for storing the subproblems, which is indexed by a vertex of $S_i$, a tree $T \in \T$, and a distribution $F$. Formally, we maintain a table of boolean entries $DP_i[u,\tilde{T},\tilde{F}]$ where $u \in S_i$, $\tilde{T}$ is a word with $O(\ell \log d)$ bits encoding a tree $T \in \T$, and $\tilde{F}$ is a word with $O(\ell \log \ell)$ bits encoding a distribution $F$. We will only use those entries of the table which correspond to the spanning tree of a connected subset $K$ with $u$ the maximum element of $K$. Thus, we say that a triple $(u,\tilde{T},\tilde{F})$ is a \emph{representative} of a subproblem $t = (i,K,F)$ if $u$ is the maximum element of $K$, $\tilde{T}$ encodes the tree $T$ such that $\Phi(N_i,u,T)$ is a spanning tree of $G_i[K]$, and $\tilde{F}$ encodes the distribution $F$. An entry of $DP_i[u,\tilde{T},\tilde{F}]$ will be called \emph{valid} if $(u,\tilde{T},\tilde{F})$ is the representative of a subproblem $t$, in which case the entry will indicate whether $t$ is satisfiable; invalid entries will be ignored.

Finally, let us explain optimization (iii). Suppose that step $i+1$ merges rectangles $j_1$ and $j_2$ into $j$. When we move to step $i+1$, we want that the valid entries $DP_i[u,\tilde{T},\tilde{F}]$ from step $i$ either keep the same value or become invalid. Thus, we need to define the adjacency functions $N_i$ in such a way that when we move to step $i+1$, a vertex $k \in S_i$ retains the same numbered neighbors in $N_{i+1}$ as in $N_i$, except possibly for $j_1,j_2$ and $j$. To ensure this, we will maintain for each vertex $k \in S_i$ and $\alpha \in \{1,2\}$, an array $\Intt_{\alpha}[k]$ of length $d$ whose entries contain elements of $S_i \cup \{\perp\}$, such that the non-null entries of $\Intt_{\alpha}[k]$ correspond to the elements of $\view_{\alpha}(\R_i,k)$. When moving to step $i+1$, we update these tables in the straightforward way: each occurrence of $j_1$ and $j_2$ is replaced with $\perp$, and if $k$ now $\alpha$-views $j$ then the first null entry is replaced by $j$. The adjacency function $N_i$ is defined from these tables at step $i$: given $k \in S_i$, (i) if $1 \leq j \leq d$ then $N_i(k,j) = \Intt_1[k][j]$, (ii) if $d+1 \leq j \leq 2d$ then $N_i(k,j) = \Intt_2[k][j-d]$. Then for each valid entry $DP_i[u,\tilde{T},\tilde{F}]$, either $\Phi(N_i,u,T)$ intersects $\{j_1,j_2\}$, in which case the entry becomes invalid in $DP_{i+1}$ (as $\Phi(N_{i+1},u,T)$ is either $\perp$ or is a graph containing $j > u$), or $\Phi(N_i,u,T)$ is disjoint from $\{j_1,j_2\}$, in which case the entry is equal to $DP_{i+1}[u,\tilde{T},\tilde{F}]$ (as the update of $\Intt_{\alpha}$ and the validity of the entry ensure that $\Phi(N_{i+1},u,T) = \Phi(N_i,u,T)$ which thus does not contain $j$). This means that we only need to compute the entries of $DP_{i+1}$ corresponding to the new rectangle $j$.

\begin{proposition} \label{p:algo-subpattern} The \textsc{Permutation Pattern} problem can be solved in time $O((16\ell d)^\ell \ell^2 n)$ if a $d$-wide decomposition of $\pi$ is given in the input.
\end{proposition}

\begin{proof} For efficiency, we represent the rectangle family $\R = (S,R)$ by the following data structure $\Delta$:
\begin{itemize}
\item a linked list $\HEndpoints$ containing pointers to the horizontal endpoints, sorted by increasing order;
\item a linked list $\VEndpoints$ containing pointers to the vertical endpoints, sorted by increasing order;
\item for each $p \in \HEndpoints \cup \VEndpoints$, an integer $\rect(p)$ equal to the index of the rectangle with endpoint $p$, and an integer $\coord(p)$ inducing an increasing numbering of the endpoints;
\item an array $\Rects$ indexed by $S$, such that $\Rects[k]$ contains a tuple $(x_1,x_2,y_1,y_2)$ representing rectangle $R(k)$, where $x_1,x_2$ point to entries of $\HEndpoints$ and $y_1,y_2$ point to entries of $\VEndpoints$;
\item for each $\alpha \in \{1,2\}$, an array $\Intt_{\alpha}$ indexed by $S$, such that for each $k \in S$, $\Intt_{\alpha}[k]$ is an array of length $d$ whose entries contain elements of $S \cup \{\perp\}$.
\end{itemize}

The data structure supports the operation \textsc{Merge}, which takes $\Delta$ representing $\R = (S,R)$, three elements $i,j \in S$ and $k$, and updates the data structure to represent $\R' = \R [i,j \rightarrow k]$. First, it sets $\Rects[k]$ to a tuple $(x_1,x_2,y_1,y_2)$ representing the new rectangle, and it removes from $\HEndpoints$ and from $\VEndpoints$ the four endpoints that are no longer present. Second, for each $\alpha \in \{1,2\}$ it updates $\Intt_{\alpha}$ as follows. For each non-null entry $v$ in $\Intt_{\alpha}[i]$, we replace the occurrence of $i$ with $\perp$ in $\Intt_{\alpha}[v]$, and for each non-null entry $v$ in $\Intt_{\alpha}[j]$ we replace the occurrence of $j$ with $\perp$ in $\Intt_{\alpha}[v]$. Next, for each $\alpha \in \{1,2\}$: (i) we construct the set $V_{\alpha} = \view_{\alpha}(\R',k)$; (ii) we define the first entries of $\Intt_{\alpha}[k]$ to contain $V_{\alpha}$ and we set the remaining entries to $\perp$; (iii) for each $v \in V_{\alpha}$ we replace the first null entry of the array $\Intt_{\alpha}[v]$ by $k$. Let us describe the construction of $V_{\alpha}$ in step (i). If $I_{\alpha}(R(i)) \subseteq I_{\alpha}(R(j))$ then $V_{\alpha}$ contains the non-null elements of $\Intt_{\alpha}[j]$; if $I_{\alpha}(R(j)) \subseteq I_{\alpha}(R(i))$ then $V_{\alpha}$ contains the non-null elements of $\Intt_{\alpha}[i]$. Suppose now that $I_{\alpha}(R(i))$ and $I_{\alpha}(R(j))$ are incomparable, and that the first is to the left of the second. We start with $V_{\alpha}$ containing the non-null elements of $\Intt_{\alpha}[i]$, then we scan the endpoints by starting from the right endpoint of $I_{\alpha}(R(i))$ and moving to the right; for each left endpoint encountered corresponding to a rectangle $v$, we add $v$ to $V_{\alpha}$; we stop when we have reached the right endpoint of $I_{\alpha}(R(j))$.

Suppose that we start for $i = 0$ with $\Delta$ representing $\R_0 = \pi$, and that for each $i$ from $1$ to $n-1$, if step $i$ merges $j_1,j_2$ into $j$ then we call $\textsc{Merge}(\Delta,j_1,j_2,j)$. We claim that the following holds at each step:
\begin{itemize}
\item[(i)] $\Delta$ represents $\R_i$;
\item[(ii)] for each $k \in S_i,\alpha \in \{1,2\}$, the non-null entries of $\Intt_{\alpha}[k]$ are the elements of $\view_{\alpha}(\R_i,k)$;
\item[(iii)] each call to \textsc{Merge} takes time $O(d^2)$.
\end{itemize}
Points (i)-(ii)-(iii) clearly hold initially. Suppose that they hold at step $i$, let us consider step $i+1$, and suppose that it merges $j_1,j_2$ into $j$. Point (i) holds by definition of $\Rects[j]$ and by induction hypothesis. Point (ii) holds for $j$ as by construction the array $\Intt_{\alpha}[j]$ contains the elements of the set $\view_{\alpha}(\R_{i+1},j)$, which has size less than $d$. Let us show that point (ii) holds for $k \neq j$. Observe that if $\view_{\alpha}(\R_{i+1},k)$ does not contain $j$ then $\Intt_{\alpha}[k]$ is unchanged, and thus its non-null entries are those of $\view_{\alpha}(\R_{i+1},k) = \view_{\alpha}(\R_i,k)$. Suppose now that $\view_{\alpha}(\R_{i+1},k)$ contains $j$. If $\view_{\alpha}(\R_i,k)$ contained $j_1$ or $j_2$, then in $\Intt_{\alpha}[k]$ we have removed $j_1$ or $j_2$ and added an occurrence of $j$. If $\view_{\alpha}(\R_i,k)$ contained neither $j_1$ nor $j_2$, then in $\Intt_{\alpha}[k]$ we have added an occurrence of $j$ (this is possible as they were at most $d-1$ non-null entries of $\Intt_{\alpha}[k]$ by (ii)). In both cases we conclude that after the update, the non-null entries of $\Intt_{\alpha}[k]$ are the elements of $\view_{\alpha}(\R_{i+1},k)$. For point (iii), observe that in a call to $\textsc{Merge}(\Delta,j_1,j_2,j)$, the construction of the set $V_{\alpha} = \view_{\alpha}(\R_{i+1},j)$ takes $O(|V_{\alpha}|) = O(d)$ time, and we need to examine $O(d)$ elements $k \in S_i$ and for each of them to update its array $\Intt_{\alpha}[k]$ in $O(d)$ time.

We now describe our main algorithm, that maintains a data structure $\Delta$ representing $\R_i$, and a table $DP_i$ as explained above. Observe that we can perform the initialization for step $i = 0$ in $O(\ell n)$ time: indeed, $\R_0$ contains the points of $\pi$, and the table $DP_0$ can be filled in a straightforward way as the only admissible subproblems have $|K| = 1$. Suppose that we have computed this information for index $i$, we show how to update the information to step $i+1$ in $O((16 \ell d)^\ell \ell^2)$ time. Suppose that step $i+1$ merges $j_1$ and $j_2$ into $j$. We can update $\Delta$ to represent $\R_{i+1}$ in $O(d^2)$ time by calling $\textsc{Merge}(\Delta,j_1,j_2,j)$, so let us focus on the update of $DP_i$. By point (ii) above, the function $N_{i+1}$ defined from $\Intt_{\alpha}$ is an adjacency function for $G_{i+1}$; thus, as explained above, the only entries of $DP_{i+1}$ differing from $DP_i$ are the valid entries $DP_{i+1}[j,\tilde{T},\tilde{F}]$. We compute these entries by enumerating each tree $T \in \T$ and each distribution $F$. There are $O((8d)^\ell)$ choices for $T$, $O(\ell^\ell)$ choices for $F$, and we claim that the computation takes $O(2^\ell \ell^2)$ time for a given choice. In $O(\ell)$ time, we first compute $G' = \Phi(N_{i+1},j,T)$, and test if $G'$ is a tree inducing a set $K$. If this does not hold, we set $DP_{i+1}[j,\tilde{T},\tilde{F}]$ to false; otherwise, the entry represents a subproblem $t = (i+1,K,F)$, and we compute $DP_{i+1}[j,\tilde{T},\tilde{F}]$ according to Lemma \ref{l:recurrence}. We need to go through the $O(2^\ell)$ partitions $\Pi$, and for a given partition $\Pi$: (i) to compute (the representatives of) the subproblems of $T^{\Pi}$ associated to $t$, (ii) to check that they are satisfiable, (iii) to check that they are pairwise compatible. Step (i) can be done by computing in $O(\ell^2)$ time a spanning forest of $G_i[K^{\Pi}]$. Step (ii) takes $O(\ell)$ time by accessing the table $DP_i$, and Step (iii) takes $O(\ell^2)$ time. Thus, this last part takes $O((16 d \ell)^\ell \ell^2)$ time.

Overall, the total running time is $O((16 d \ell)^\ell \ell^2 n)$, and we decide if $\sigma$ is a subpattern of $\pi$ by checking at the last step if the subproblem $(n-1,\{j\},F)$ is satisfiable, where $j$ is the unique rectangle of $\R_{n-1}$ and $F(j) = S(\sigma)$. 
\end{proof}  
\fi

\section{Hardness results}\label{sec:hardness-results}

In this section, we establish the $\wone$-hardness of some variants of the \textsc{Permutation Pattern} problem. We first consider the following constrained version of the problem.

\prob{\textsc{Partitioned Permutation Pattern}}{Two permutations $\sigma$ and $\pi$, a partition of $S(\pi)$ in sets $S_i$ ($i \in S(\sigma)$)}{Does there exist an embedding $\phi$ of $\sigma$ into $\pi$ such that $\phi(i) \in S_i$ for each $i \in S(\sigma)$?}

\begin{theorem} \label{t:hardness-colored} \textsc{Partitioned Permutation Pattern} is $\wone$-hard for parameter $|\sigma|$, even when $\sigma$ is a canonical $r \times r$-grid.
\end{theorem}
\iffull
\begin{proof} We give a reduction from \textsc{Partitioned Clique} \cite{DBLP:journals/jcss/Pietrzak03}. Let $\I$ be an instance of \textsc{Partitioned Clique}, consisting of a graph $H = (V,E)$, an integer $k$, and a partition of $V$ into sets $V_1,\ldots,V_k$. We let $\sigma$ be the canonical $(3k+1) \times (3k+1)$-grid. We now describe the construction of $\pi$. For each $i \in [k]$, let $v^1_i,\ldots,v^{n_i}_i$ be an enumeration of $V_i$. We start with a rectangle $R = I \times J$, and we subdivide $I$ into intervals $I_0,\ldots,I_k$ and $J$ into intervals $J_0,\ldots,J_k$. Then we subdivide each interval $I_i$ ($i \in [k]$) into $3 n_i$ consecutive intervals $I^1_{i,j},I^2_{i,j},I^3_{i,j}$ ($1 \leq j \leq n_i$), and we subdivide each interval $J_i$ ($i \in [k]$) into $3 n_i$ consecutive intervals $J^1_{i,j},J^2_{i,j},J^3_{i,j}$ ($1 \leq j \leq n_i$). Let $G$ be the resulting $(3n+1) \times (3n+1)$-gridding of $R$. Each cell $C$ of $G$ will contain at most one point of $\pi$ according to the following criterion: (i) if $C = I^2_{i,x} \times J^2_{j,y}$, then $C$ contains a point iff ($i = j$ and $x = y$) or ($i \neq j$ and $\{v^i_x,v^j_y\} \in E$); (ii) every other cell contains one point. It remains to describe the horizontal ordering of the points inside a column of $G$, and the vertical ordering of the points inside a row of $G$. Inside a column $x$ of $G$ corresponding to an interval $I^r_{i,j}$, we order the points such that if $p \in G(x,y)$ and $p' \in G(x,y')$, then $\pr_1(p) < \pr_1(p')$ iff $y' < y$. Inside a row $y$ of $G$ corresponding to an interval $J^r_{i,j}$, we order the points such that if $p \in G(x,y)$ and $p' \in G(x',y)$ then $\pr_2(p) < \pr_2(p')$ iff $x < x'$. Inside column 1 of $G$ corresponding to the interval $I_0$, we order the points such that if $p \in G(1,y)$ and $p' \in G(1,y')$, then: (i) if $y$ corresponds to $J^r_{i,j}$ and $y'$ corresponds to $J^s_{i,j'}$ with $j < j'$, then $\pr_1(p) < \pr_1(p')$; (ii) in all other cases, $\pr_1(p) < \pr_1(p')$ iff $y' < y$. Inside row 1 of $G$ corresponding to the interval $J_0$, we order the points such that if $p \in G(x,1)$ and $p' \in G(x',1)$, then: (i) if $x$ corresponds to $I^r_{i,j}$ and $x'$ corresponds to $I^s_{i,j'}$ with $j' < j$, then $\pr_2(p) < \pr_2(p')$; (ii) in all other cases, $\pr_2(p) < \pr_2(p')$ iff $x < x'$. We let $\pi$ be the resulting permutation. Finally, we define the sets $S_i$ ($i \in S(\sigma)$) as follows. Let us denote by $p(x,y)$ the element of $S(\sigma)$ corresponding to the point in the $(x,y)$th cell of the gridding of $\sigma$. For a point $p$ of $\pi$, we define $f_1(p)$ such that if $p \in I_0 \times J$ then $f_1(p) = 1$, and if $p \in I^r_{i,j} \times J$ then $f_1(p) = 3(i-1)+r+1$; we define $f_2(p)$ symmetrically, and we put the point $p$ in $S_{p(f_1(p),f_2(p))}$. Let $\I'$ be the resulting instance of \textsc{Partitioned Permutation Pattern}, then $\I'$ can clearly be constructed in polynomial time.

We now argue for the correctness of the reduction. Suppose that $H$ has a clique $C = \{v^{p_1}_1,\ldots,v^{p_k}_k\}$. Given $2 \leq i \leq 3k+1$, let $f(i) = ((i-2) $ div $ 3 + 1, (i-2) $ mod $ 3 + 1)$. We define $\phi : S(\sigma) \rightarrow S(\pi)$ as follows:
\begin{itemize}
\item We map $p(1,1)$ to the unique point of $\pi$ in $I_0 \times J_0$;
\item for $2 \leq x \leq 3k+1$, let $(i,r) = f(x)$, then we map $p(x,1)$ to the unique point of $\pi$ in $I^r_{i,p_i} \times J_0$;
\item for $2 \leq y \leq 3k+1$, let $(j,s) = f(y)$, then we map $p(1,y)$ to the unique point of $\pi$ in $I_0 \times J^s_{j,p_j}$;
\item for $2 \leq x,y \leq 3k+1$, let $(i,r) = f(x)$ and $(j,s) = f(y)$, then we map $p(x,y)$ to the unique point of $\pi$ in $I^r_{i,p_i} \times J^s_{j,p_j}$.
\end{itemize}
Note that in the last case, the existence of the point follows from the fact that $\{v^i_{p_i},v^j_{p_j}\} \in E$. We then have $\phi(p) \in S_p$ for each $p \in S(\sigma)$, and it can be checked that $\phi$ is an embedding of $\sigma$ into $\pi$. Conversely, suppose that $\phi$ is an embedding of $\sigma$ into $\pi$ such that $\phi(p) \in S_p$ for each $p \in S(\sigma)$. Given $i \in [k]$, consider the elements $q^1_i = p(3(i-1)+2,1), q^2_i = p(3(i-1)+3,1), q^3_i = p(3(i-1)+4,1)$ in $S(\sigma)$. We have $q^1_i <^{\sigma}_2 q^2_i <^{\sigma}_2 q^3_i$, and by the arrangement of the points it means that there exists $1 \leq p_i \leq n_i$ such that $\phi(q^a_i)$ is the unique point of $I^a_{i,p_i} \times J_0$. Likewise, given $j \in [k]$, by considering the points $r^1_j = p(1,3(j-1)+2), r^2_j = p(1,3(j-1)+3), r^3_j = p(1,3(j-1)+4)$, we obtain that there exists $1 \leq p'_j \leq n_j$ such that $\phi(r^a_j)$ is the unique point of $I_0 \times J^a_{j,p'_j}$. Now, for $i,j \in [k]$, consider $s_{i,j} = (3(i-1)+3,3(j-1)+3)$. Since $q^1_i <^{\sigma}_1 s_{i,j} <^{\sigma}_1 q^3_i$, it follows that $\phi(s_{i,j})$ is in $I^2_{i,p_i} \times J$; since $r^1_j <^{\sigma}_2 s_{i,j} <^{\sigma}_2 r^3_j$, it follows that $\phi(s_{i,j})$ is in $I \times J^2_{j,p'_j}$; thus $\phi(s_{i,j})$ is the unique point of $I^2_{i,p_i} \times J^2_{j,p'_j}$. In particular, since for $i \in [k]$ the point $\phi(r_{i,i})$ is present we obtain that $p_i = p'_i$, and since for $i,j \in [k]$ distinct the point $\phi(r_{i,j})$ is present we obtain that $\{v^{p_i}_i,v^{p_j}_j\} \in E$. We conclude that $C = \{ v^{p_1}_1, \ldots, v^{p_k}_k\}$ is a clique of $H$.
\end{proof}
\fi

We now consider the generalization of the \textsc{Permutation Pattern} problem to $d$-dimensional permutations. Given an integer $d$, a \emph{$d$-dimensional point} is a tuple $p = (x_1,\ldots,x_d) \in \nat^d$, and for $\alpha \in [d]$ we define $\pr_{\alpha}(p) = x_{\alpha}$. A \emph{$d$-dimensional permutation} is defined as a pair $\pi = (S,P)$ with $S$ a set and $P : S \rightarrow \nat^d$ an injection such that $P(S)$ is a set of $d$-dimensional points in general position (i.e. for each $\alpha \in [d]$ it holds that $\pr_{\alpha}$ is injective on $P(S)$); we let $S(\pi) = S$. Given $p,p' \in S$ and $\alpha \in [d]$, we denote $p <^{\pi}_{\alpha} p'$ iff $\pr_{\alpha}(P(p)) < \pr_{\alpha}(P(p'))$. Given two $d$-dimensional permutations $\sigma$ and $\pi$, an \emph{embedding} of $\sigma$ into $\pi$ is a function $\phi : S(\sigma) \rightarrow S(\pi)$ such that for every $p,p' \in S(\sigma)$, for every $\alpha \in [d]$, $p <^{\sigma}_{\alpha} p'$ iff $\phi(p) <^{\pi}_{\alpha} \phi(p')$. We consider the following problem.\\

\prob{\textsc{$d$-Dimensional Permutation Pattern}}{Two $d$-dimensional permutations $\sigma$ and $\pi$.}{Does there exist an embedding of $\sigma$ into $\pi$?}

\begin{theorem} \label{t:hardness-ddim} For every $d \geq 3$, \textsc{$d$-Dimensional Permutation Pattern} is $\wone$-hard for parameter $|\sigma|$.
\end{theorem}
\iffull
\begin{proof} We prove the result for $d = 3$, since the extension to any $d \geq 3$ is straightforward. We give the following reduction from \textsc{Partitioned Permutation Pattern}. Let $\I$ be an instance of \textsc{Partitioned Permutation Pattern}, consisting of a permutation $\sigma$ with $S(\sigma) = [\ell]$, a permutation $\pi$, and a partition of $S(\pi)$ in sets $S_1,\ldots,S_{\ell}$. We assume w.l.o.g. that for $i,j \in [\ell]$, $i < j$ iff $i <^{\sigma}_1 j$. Suppose that $\sigma = (S_{\sigma},P_{\sigma})$ and $\pi = (S_{\pi},P_{\pi})$. We define two 3-dimensional permutations $\sigma' = (S_{\sigma},P_{\sigma'})$ and $\pi' = (S_{\pi},P_{\pi'})$ as follows. For each $i \in S_{\sigma}$, if $P_{\sigma}(i) = (x,y)$ then we set $P_{\sigma'}(i) = (x,y,i)$. Now, we define a numbering $f(i)$ of the points of $\pi$ as follows: we first number the points of $S_1$ by decreasing $x$-coordinate, then the points of $S_2$ by decreasing $x$-coordinate, etc. For each $i \in S_{\pi}$, if $P_{\pi}(i) = (x,y)$ then we set $P_{\pi'}(i) = (x,y,f(i))$. Let $\I' = (\sigma',\pi')$ be the resulting instance, observe that $\I'$ can be constructed in polynomial time. We show that $\I$ is a positive instance of \textsc{Partitioned Permutation Pattern} iff $\I'$ is a positive instance of \textsc{3-Dimensional Permutation Pattern}.

Suppose that $\I$ is a positive instance via an embedding $\phi$ of $\sigma$ into $\pi$. Given $i,j \in S_{\sigma}$ distinct, for each $\alpha \in \{1,2\}$ we have that $i <^{\sigma'}_{\alpha} j$ iff $\phi(i) <^{\pi'}_{\alpha} \phi(j)$ (by definition of $\phi$), and for $\alpha = 3$ we have that $i <^{\sigma'}_3 j$ iff $i < j$ iff $f(\phi(i)) < f(\phi(j))$ (since $\phi(i) \in S_i$ and $\phi(j) \in S_j$, and as $i,j$ are distinct) iff $\phi(i) <^{\pi'}_3 \phi(j)$. Conversely, suppose that $\I'$ is a positive instance via an embedding $\phi$ of $\sigma'$ into $\pi'$. Clearly, $\phi$ is also an embedding of $\sigma$ into $\pi$, and there is a function $\psi : [\ell] \rightarrow [\ell]$ such that $\phi(i) \in S_{\psi(i)}$ for each $i \in [\ell]$. Suppose by contradiction that $\psi$ is not the identity function, then there exist $i,j \in [\ell]$ such that $i < j$ and $\psi(j) \leq \psi(i)$. If $\psi(i) = \psi(j)$, we obtain that $i <^{\sigma}_1 j$ and thus $\phi(i) <^{\pi}_1 \phi(j)$, but then $f(\phi(j)) < f(\phi(i))$; if $\psi(j) < \psi(i)$, we also obtain that $f(\phi(j)) < f(\phi(i))$. In both cases, we obtain that $\phi(j) <^{\pi'}_3 \phi(i)$ and $i <^{\sigma'}_3 j$, contradicting the assumption that $\phi$ is an embedding.
\end{proof}
\fi
\newcommand{\mymid}{\textup{mid}}
\newcommand{\pin}{\textup{pin}}
\newcommand{\cont}{\textup{cont}}

\section{The case of $t$-monotone permutations}\label{sec:t-monotone}

The notions of \emph{increasing} and \emph{decreasing}  permutations are defined  the obvious way and we will 
use \emph{monotone} for a permutation that is either increasing or decreasing. Formally, let $\pi$ be a permutation, we say that $\pi$ is \emph{increasing} (resp.,~\emph{decreasing}) if for any $p,p' \in S(\pi)$, it holds that $p <^{\pi}_1 p'$ iff $p <^{\pi}_2 p'$ (resp.,~$p' <^P_2 p$); we say that $\pi$ is \emph{monotone} if it is either increasing or decreasing. Given an integer $t$, we say that $\pi$ is \emph{$t$-increasing} (resp.,~\emph{$t$-monotone}) if there is a partition $\Pi = (S_1,\ldots,S_t)$ of $S(\pi)$ such that $\pi | S_i$ is increasing (resp., monotone) for each $i \in [t]$. The partition $\Pi$ will be called a \emph{$t$-increasing (resp., $t$-monotone) partition}.

Let us briefly discuss the recognition problem for these classes. While $t$-increasing permutations can be recognized in polynomial time, recognizing $t$-monotone permutations is $\np$-hard for unbounded $t$ \cite{DBLP:journals/eik/BrandstadtK86}. For a fixed $t$, recoginizing $t$-monotone permutations is fixed-parameter tractable: the algorithm of Heggernes et al.~solves the problem in time $2^{O(t^2 \log t)}\cdot n^{O(1)}$ \cite{DBLP:conf/swat/HeggernesKLRS10}. We can also give a constant-factor approximation for the problem in the sense that, given a permutation $\pi$ of length $n$,  in time $O(n^2)$ we can either find a $c t$-monotone partition of $\pi$, or conclude that $\pi$ is not $t$-monotone. This easily follows from Greene theorem with $c = 2$ \cite{Greene74}, and there exists a better algorithm that yields $c = 1.71$ \cite{DBLP:journals/ipl/FominKN02}.

\iffull\subsection{Width of $t$-monotone permutations}\label{sec:width-t-monotone}\fi

It can be seen that a $t$-increasing permutation cannot have a $(t+1) \times (t+1)$-grid, and that a $t$-monotone permutation cannot have a $(2t+1) \times (2t+1)$-grid. It follows that these permutations have bounded width (at most $4 f(2t+1)$) by Theorem~\ref{t:algo-decomp}. The following result gives a better bound.

\begin{proposition} \label{p:monotone-bound} If $\pi$ is a $t$-monotone permutation, then $w(\pi) \leq 6t-5$.
\end{proposition}
\iffull
\begin{proof} We first need some additional definitions on rectangle families. Let $\R = (S,R)$ be a rectangle family. Given $S' \subseteq S$, we let $\R | S' = (S', R | S')$. We say that $\R$ is \emph{increasing} if there is an enumeration $i_1,\ldots,i_n$ of $S$ such that for each $p < q$, $I_1(R(i_p)) < I_1(R(i_q))$ and $I_2(R(i_p)) < I_2(R(i_q))$. Likewise, we say that $\R$ is \emph{decreasing} if there is an enumeration $i_1,\ldots,i_n$ of $S$ such that for each $p < q$, $I_1(R(i_p)) < I_1(R(i_q))$ and $I_2(R(i_p)) > I_2(R(i_q))$. In each case, we call \emph{consecutive} two indices of the form $i_p, i_{p+1}$. We say that $\R$ is \emph{monotone} if it is increasing or decreasing; we say that $\R$ is \emph{$t$-monotone} if it admits a $t$-monotone partition, i.e. a partition of $S$ in sets $S_1,\ldots,S_t$ such that each $\R | S_r$ is monotone.

We are now ready to prove the proposition. Suppose that $\pi$ has a $t$-monotone partition $\Pi = (S_1,\ldots,S_t)$. Starting with $\R = \pi$, we will do a sequence of merges maintaining the following invariants: (i) $\R$ is a $(6t-5)$-wide rectangle family; (ii) $\Pi$ is a $t$-monotone partition of $\R$. At each step, we proceed as follows. Suppose that $\R = (S,R)$ and $\Pi = (S_1,\ldots,S_t)$. If each set $S_r$ is a singleton, then $\R$ has at most $t$ rectangles and we can easily complete the sequence of merges. Suppose now that some set $S_r$ is not a singleton. We define the set $\M$ of \emph{mergeable pairs} as the set of pairs $(i,j)$ coming from a same set $S_r$ and that are consecutive in $\R | S_r$; observe that $\M$ is not empty. Given a pair $m = (i,j) \in \M$ coming from a set $S_r$, by \emph{merging} the pair $m$, we mean the following: (i) replace $\R$ by $\R' = \R [i,j \rightarrow k]$ where $k$ is a new index; (ii) replace $\Pi$ by $\Pi' = (S'_1,\ldots,S'_t)$ where $S'_r = S_r - \{i,j\} + \{k\}$, and $S'_s = S_s$ for $s \neq r$. Observe that after this operation, $\Pi'$ is still a $t$-monotone partition of $\R'$. We will show that we can find a mergeable pair in $\M$ whose merging results in a new rectangle $k$ with $\view(\R',k) < 6t-5$.

Consider a pair $m = (i,j) \in \M$, and let $R$ be the smallest rectangle enclosing $R(i) \cup R(j)$. For each $\alpha \in \{1,2\}$, we define $\pin_{\alpha}(m)$ as the set of elements $i' \in S - \{i,j\}$ such that $I_{\alpha}(R(i')) \subseteq I_{\alpha}(R)$, we define the sum $\Sigma_{\alpha} := \sum_{m \in \M} |\pin_{\alpha}(m)|$, and we define $\Sigma := \sum_{m \in \M} \max(|\pin_1(m)|,|\pin_2(m)|)$.\\

\begin{claim}\label{cl:mon1} For each $\alpha \in \{1,2\}$, $\Sigma_{\alpha} \leq 2(t-1) |\M|$.
\end{claim}
\begin{proof} Fix $\alpha \in \{1,2\}$. We say that an element $i \in S$ \emph{contributes} to a pair $m \in \M$ if $i \in \pin_{\alpha}(m)$; we let $\cont(i)$ denote the number of pairs $m \in \M$ to which $i$ contributes. Then clearly $\Sigma_{\alpha} = \sum_{i \in S} \cont(i)$. Observe that an element $i \in S_r$ contributes to no pair in $S_r$, and to at most two pairs in each set $S_s$ ($s \neq r$), hence $\cont(i) \leq 2(t-1)$. This yields that $\Sigma_{\alpha} \leq 2 (t-1) |S|$. We can slightly improve the bound to $2 (t-1) |\M|$ as follows. For each $s \in [t]$, let $i_s,i'_s$ be the first and last indices in the natural enumeration of $S_s$. Let us sort the elements $s \in [t]$ by increasing order of the left endpoint of $I_{\alpha}(R(i_s))$; this gives an enumeration $E_1$ of $[t]$. Likewise, let us sort the elements $s \in [t]$ by decreasing order of the right endpoint of $I_{\alpha}(R(i'_s))$; this gives an enumeration $E_2$ of $[t]$. Now, if $s$ is the $p$th element of $E_1$ (resp., $E_2$), we have that $\cont(i_s) \leq 2(p-1)$ (resp., $\cont(i'_s) \leq 2(p-1)$). It follows that $\Sigma_{\alpha} \leq 2 (t-1) (|S| - 2t) + 2 \sum_{p = 1}^{t} 2 (p-1) = 2 (t-1) |S| - 4t(t-1) + 2 t (t-1) = 2 (t-1) (|S|-t) = 2 (t-1) |\M|$.
\cqed\end{proof}
Now, for each $m \in \M$, we have $\max(|\pin_1(m)|,|\pin_2(m)|) \leq |\pin_1(m)| + |\pin_2(m)|$, and thus $\Sigma \leq \sum_{m \in \M} (|\pin_1(m)| + |\pin_2(m)|) = \Sigma_1 + \Sigma_2 \leq 4(t-1) |\M|$. Hence, we can find a pair $m \in \M$ coming from a set $S_r$ such that for each $\alpha \in \{1,2\}$ it holds that $|\pin_{\alpha}(m)| \leq 4(t-1)$. Consider the result of merging pair $m$ into a new rectangle $k$, thus yielding the rectangle family $\R'$ and the $t$-monotone partition $\Pi' = (S'_1,\ldots,S'_t)$.

\begin{claim}\label{cl:mon2} $\view(\R',k) \leq 6(t-1)$.
\end{claim}
\begin{proof} We show that $|\view_{\alpha}(\R',k)| \leq 6(t-1)$ holds for each $\alpha \in \{1,2\}$. Let $V = \view_{\alpha}(\R',k)$, we partition $V$ in two sets $V_1 := \pin_{\alpha}(m)$ and $V_2 := V \backslash V_1$. Observe that $S'_r$ contains no element from $V$, and that for $s \neq r$ the set $S'_s$ can contain at most two elements from $V_2$ (for if $S'_s$ contains three elements $u,v,w \in V$ with $I_{\alpha}(R(u)) < I_{\alpha}(R(v)) < I_{\alpha}(R(w))$, then $v \in V_1$). It follows that $|V_2| \leq 2(t-1)$, and as we also have $|V_1| = |\pin_{\alpha}(m)| \leq 4(t-1)$, we conclude that $|V| \leq 6(t-1)$ as claimed.
\cqed\end{proof}
That is, $\R'$ is also $(6t-5)$-wide, as required.
\end{proof}
\fi

\iffull Observe that the \fi proof of Proposition~\ref{p:monotone-bound} can be turned into an algorithm that takes a permutation $\pi$ of length $n$ together with a $t$-monotone partition, and produces in time $O(n^2)$ a $(6t-5)$-wide decomposition of $\pi$. Combining this with Theorem \ref{th:dp}, this yields a $t^{O(\ell)} n^2$ algorithm for the \textsc{Permutation Pattern} problem on $t$-monotone permutations. However, it turns out that this problem admits a very simple algorithm using the theory of constraint satisfaction problems and completely independent of our decomposition and width measure\iffull; we present this algorithm in the next section\fi.

\iffull\subsection{CSPs and $t$-monotone permutations}\label{sec:csps-t-monotone}\fi
We present an algorithm for solving \textsc{Permutation Pattern} on
$t$-monotone instances by reducing it to a constraint satisfaction
problem. The algorithm relies on the known fact that a CSP instance
with a majority polymorphism can be solved in polynomial time.

\iffull
As our use of CSP techniques is standard and what makes it surprising is the observation is that these techniques solve the problem immediately, we recall only briefly the most important notions related to CSPs. For more background, the reader is referred to, e.g., the survey \cite{DBLP:journals/csur/Chen09}. 
\begin{definition}\label{def:csp}
An instance of a {\em constraint satisfaction problem} is a triple $(V ,D, C)$,
where:
\begin{itemize}[noitemsep]
\item $V$ is a set of variables,
\item $D$ is a domain of values,
\item $C$ is a set of constraints, $\{c_1,c_2,\dots ,c_q\}$.
Each constraint $c_i\in C$ is a pair $\langle
s_i,R_i\rangle$, where:
\begin{itemize}[noitemsep]
\item $s_i$ is a tuple of variables of length $m_i$, called the {\em constraint scope,} and
\item $R_i$ is an $m_i$-ary relation over $D$, called the {\em constraint
  relation.}
\end{itemize}
\end{itemize}
\end{definition}
For each constraint $\langle s_i,R_i\rangle$ the tuples of $R_i$
indicate the allowed combinations of simultaneous values for the
variables in $s_i$. The length $m_i$ of the tuple $s_i$ is called the
{\em arity} of the constraint.  A {\em solution} to a constraint
satisfaction problem instance is a function $f$ from the set of
variables $V$ to the domain of values $D$ such that for each
constraint $\langle s_i,R_i\rangle$ with $s_i = \langle
v_{i_1},v_{i_2},\dots,v_{i_m}\rangle$, the tuple $\langle f(v_{i_1}),
f(v_{i_2}),\dots,f(v_{i_m})\rangle$ is a member of $R_i$.

A \emph{polymorphism} of a (say, $n$-ary) relation $R$ on $D$ is a
mapping $f:D^k\to D$ for some $k$ such that for any tuples $\mathbf{a}_1,
\ldots,\mathbf{a}_k\in R$ the tuple  
\[
f(\mathbf{a}_1,\ldots,\mathbf{a}_k)=(f(\mathbf{a}_1[1],\dots,\mathbf{a}_k[1]),\dots,f(\mathbf{a}_1[n],\dots,\mathbf{a}_k[n]))
\]
belongs to $R$. A {\em majority polymorphism} is a ternary
polymorphism $f$ with the property that $f(x,x,y)=f(x,y,x)=f(y,x,x)=x$
for any $x,y\in D$. It is known that if there is a function $f$ that
is a majority polymorphism for {\em every} constraint of the instance,
then the instance can be solved in polynomial time
\cite{DBLP:journals/siamcomp/FederV98}.

We solve a constrained version of the \textsc{Permutation Pattern} problem, where the image of each
element of $\sigma$ has to be in a prespecified monotone sequence. That
is, given two permutations $\sigma$ and $\pi$, given a $t$-monotone
partition $\Sigma = (S_1,\ldots,S_t)$ of $\sigma$ and a $t$-monotone
partition $\Pi = (S'_1,\ldots,S'_t)$ of $\pi$, the task is to find
an embedding $\phi$ of $\sigma$ into $\pi$ such that $\phi(S_i)
\subseteq S'_i$ holds for each $i \in [t]$. Such an embedding $\phi$
will be called a $(\Sigma,\Pi)$-embedding. We show that the
$(\Sigma,\Pi)$-embedding problem is polynomial-time solvable; then by
trying all possible partitions $\Sigma$, we get an algorithm for the
original \textsc{Permutation Pattern} problem on $t$-monotone
permutations.

We define a CSP instance $I=(V,D,C)$ with $V=S(\sigma)$ and
$D=S(\pi)$. The intended meaning of the value of variable $x\in
S(\sigma)$ is the image of $x$ in the embedding, or in other words, we
want to introduce constraints such that there is a one-to-one
correspondence between the solutions of $I$ and the
$(\Sigma,\Pi)$-embeddings. The constraints are defined as follows. For
each $x,y\in S(\sigma)$ and $\alpha\in\{1,2\}$, if $x <^\sigma_\alpha y$
holds, then we introduce the constraint $\langle (x,y),
R_{x,y,\alpha}\rangle$, where $R_{x,y,\alpha}$ is defined as follows. Suppose that $x\in S_i$ and $y\in S_j$ (possibly $i=j$); we let
\[
R_{x,y,\alpha}=\{ (x',y') \mid x'\in S'_i, y'\in S'_j, x' <^\pi_\alpha y' \}.
\] 
That is, the images of $x$ and $y$ have to appear in the prespecified
classes of the partition and have to respect the same ordering
relation in $\pi$ as in $\sigma$. It is easy to see that indeed there
is a correspondence between solutions and embeddings.

Our goal is to show that there is a function $f$ that is majority polymorphism for every constraint in $I$. Given three elements $x'_1,x'_2,x'_3\in S(\pi)$ and an $\alpha \in \{1,2\}$, we define $\mymid_\alpha(x'_1,x'_2,x'_3)$ as the median value $x'$ of these three elements with respect to the ordering $\le^\pi_\alpha$, that is, at most one of $\{x'_1,x'_2,x'_3\}$ is strictly larger than $x'$ and at most one element is strictly smaller than $x'$; note that this value $x'$ is well defined. The crucial observation where monotone sequences come into play is that if $x'_1,x'_2,x'_3\in S'_i$, i.e., they come from the same monotone sequence, then $\mymid_1(x'_1,x'_2,x'_3)=\mymid_2(x'_1,x'_2,x'_3)$. This allows us to show that both of these functions are polymorphisms of every constraint:
\begin{proposition}\label{p:csp-mid}
Both $\mymid_1$ and $\mymid_2$ are polymorphisms of every constraint in $I$.
\end{proposition}
\begin{proof}
  Suppose that $(x'_1,y'_1),(x'_2,y'_2),(x'_3,y'_3)\in
  R_{x,y,\alpha}$; we need to show that
\begin{gather*}
(\mymid_1(x'_1,x'_2,x'_3),\mymid_1(y'_1,y'_2,y'_3))\in R_{x,y,\alpha}\\
(\mymid_2(x'_1,x'_2,x'_3),\mymid_2(y'_1,y'_2,y'_3))\in R_{x,y,\alpha}
\end{gather*}
Suppose that $x\in S_i$ and $y\in S_j$ hold; it
  follows that $x'_1,x'_2,x'_3\in S'_i$ and $y'_1,y'_2,y'_3\in
  S'_j$. Therefore, as observed above, $\mymid_1$ and $\mymid_2$
  coincide on these values, thus it is sufficent to prove the
  statement for $\mymid_\alpha$.

  It is clear that $\mymid_\alpha(x'_1,x'_2,x'_3)\in S'_i$ and
  $\mymid_\alpha(y'_1,y'_2,y'_3)\in S'_j$. Thus we need to show only
  $\mymid_\alpha(x'_1,x'_2,x'_3)\le^\pi_\alpha
  \mymid_\alpha(y'_1,y'_2,y'_3)$. This is simply the well-known fact that
  the median function is a polymorphism of a linear ordering. For completeness, we provide a simple proof. Without loss of generality, suppose that
$x'_1 \le^\pi_\alpha x'_2 \le^\pi_\alpha x'_3$, that is, $\mymid_\alpha(x'_1,x'_2,x'_3)=x'_2$. We consider the following cases:
\begin{itemize}
\item If $\mymid_\alpha(y'_1,y'_2,y'_3)=y'_1$, then either $y'_2\le^\pi_\alpha y'_1$ (implying $x'_2\le^\pi_\alpha y'_2\le^\pi_\alpha y'_1$) or 
$y'_3\le^\pi_\alpha y'_1$ (implying $x'_2\le^\pi_\alpha x'_3 \le^\pi_\alpha y'_3 \le^\pi_\alpha y'_1$).
\item If $\mymid_\alpha(y'_1,y'_2,y'_3)=y'_2$, then  $x'_2\le^\pi_\alpha y'_2$ holds.
\item If $\mymid_\alpha(y'_1,y'_2,y'_3)=y'_3$, then  $x'_2\le^\pi_\alpha x'_3 \le^\pi_\alpha y'_3$ holds.
\end{itemize}
In all cases, we have shown that
$\mymid_\alpha(x'_1,x'_2,x'_3)\le^\pi_\alpha
\mymid_\alpha(y'_1,y'_2,y'_3)$, completing the proof.
\end{proof}

Combining Proposition \ref{p:csp-mid} with the result of \cite{DBLP:journals/siamcomp/FederV98}, we obtain a polynomial-time algorithm for the above CSP instance. Actually, we may observe that this particular CSP can be directly reduced to a 2SAT instance and can be solved in time  $O(\ell^2 n^2)$. Note that this immediately implies a fixed-parameter algorithm for the \textsc{Permutation Pattern} problem on $t$-monotone permutations: given a pattern $\sigma$, and a $t$-monotone target $\pi$ with a $t$-monotone partition $\Pi = (S_1,\ldots,S_t)$, we enumerate each possible partition $\Sigma$ of $\sigma$ into $t$ classes, test whether $\Sigma$ is a $t$-monotone partition of $\sigma$, and if so test in $O(\ell^2 n^2)$ the existence of a $(\Sigma,\Pi)$-embedding. Thus, we obtain:
\fi
\begin{theorem} \label{t:algo-monotone} Given an instance
  $(\sigma,\pi)$ of the \textsc{Permutation Pattern} problem, and a
  $t$-monotone partition $\Pi$ of $\pi$, we can solve $(\sigma,\pi)$
  in time $O(t^{\ell} \ell^2 n^2)$ and polynomial space.
\end{theorem}

We make two remarks about this result. First, it extends a result of \cite{DBLP:conf/isaac/Guillemot11} that solves the problem in $O(t^{\ell} \ell n)$ time for $t$-increasing permutations. Second, note that it assumes that a $t$-monotone partition of $\pi$ is given as input. However, if we have a promise that $\pi$ is $t$-monotone without knowing the explicit partition, then we can first obtain a $2t$-monotone partition in $O(n^2)$ time as mentioned above, and thus we can solve the problem in $O((2t)^{\ell} \ell^2 n^2)$ time for $t$-monotone permutations.

The previous theorem has an interesting consequence. Observe that a permutation of length $n$ is always $t$-monotone for $t = 2 \lceil \sqrt n \rceil$ (this can be deduced from Greene's theorem \cite{Greene74} or from Erd\H os-Szekeres theorem \cite{ErdosS35}). Plugging this into Theorem~\ref{t:algo-monotone} yields a nontrivial $n^{\frac{\ell}{2}+o(\ell)}$ time algorithm for \textsc{Permutation Pattern} using polynomial space. This has to be compared with the algorithm of \cite{DBLP:journals/siamdm/AhalR08} that uses $n^{0.47 \ell+o(\ell)}$ time and exponential space. Note also that our FPT algorithm for \textsc{Permutation Pattern} uses exponential space, due to the dynamic-programming step.

\begin{theorem} \label{t:poly-space} We can solve the \textsc{Permutation Pattern} problem in time $n^{\frac{\ell}{2}+o(\ell)}$ and polynomial space.
\end{theorem}

\bibliographystyle{abbrv}
\bibliography{permutation}

\begin{thebibliography}{10}

\bibitem{DBLP:journals/siamdm/AhalR08}
S.~Ahal and Y.~Rabinovich.
\newblock {On Complexity of the Subpattern Problem}.
\newblock {\em SIAM J. Discrete Math.}, 22(2):629--649, 2008.

\bibitem{AA05}
M.~Albert and M.~Atkinson.
\newblock {Simple permutations and pattern restricted permutations}.
\newblock {\em Discrete Math.}, 300(1--3):1--15, 2005.

\bibitem{DBLP:journals/jal/Bodlaender93}
H.~L. Bodlaender.
\newblock {On Linear Time Minor Tests with Depth-First Search}.
\newblock {\em J. Algorithms}, 14(1):1--23, 1993.

\bibitem{DBLP:journals/ijfcs/Bodlaender94}
H.~L. Bodlaender.
\newblock {On Disjoint Cycles}.
\newblock {\em Int. J. Found. Comput. Sci.}, 5(1):59--68, 1994.

\bibitem{DBLP:journals/ipl/BoseBL98}
P.~Bose, J.~F. Buss, and A.~Lubiw.
\newblock {Pattern Matching for Permutations}.
\newblock {\em Inf. Process. Lett.}, 65(5):277--283, 1998.

\bibitem{DBLP:journals/eik/BrandstadtK86}
A.~Brandst{\"a}dt and D.~Kratsch.
\newblock {On Partitions of Permutations into Increasing and Decreasing
  Subsequences}.
\newblock {\em Elektronische Informationsverarbeitung und Kybernetik},
  22(5/6):263--273, 1986.

\bibitem{DBLP:journals/corr/abs-1301-0340}
M.-L. Bruner and M.~Lackner.
\newblock {The computational landscape of permutation patterns}.
\newblock {\em CoRR}, abs/1301.0340, 2013.

\bibitem{DBLP:journals/csur/Chen09}
H.~Chen.
\newblock {A rendezvous of logic, complexity, and algebra}.
\newblock {\em ACM Comput. Surv.}, 42(1), 2009.

\bibitem{MR2465405}
A.~Claesson and S.~Kitaev.
\newblock {Classification of bijections between 321- and 132-avoiding
  permutations}.
\newblock {\em S\'em. Lothar. Combin.}, 60:Art. B60d, 30, 2008/09.

\bibitem{DBLP:books/daglib/0023376}
T.~H. Cormen, C.~E. Leiserson, R.~L. Rivest, and C.~Stein.
\newblock {\em {Introduction to Algorithms (3. ed.)}}.
\newblock MIT Press, 2009.

\bibitem{DBLP:journals/jacm/DemaineFHT05}
E.~D. Demaine, F.~V. Fomin, M.~T. Hajiaghayi, and D.~M. Thilikos.
\newblock {Subexponential parameterized algorithms on bounded-genus graphs and
  {\it H}-minor-free graphs}.
\newblock {\em J. ACM}, 52(6):866--893, 2005.

\bibitem{MR2001b:68042}
R.~G. Downey and M.~R. Fellows.
\newblock {\em {Parameterized Complexity}}.
\newblock Monographs in Computer Science. Springer, New York, 1999.

\bibitem{ErdosS35}
P.~Erd\"{o}s and G.~Szekeres.
\newblock {A combinatorial problem in geometry}.
\newblock {\em Compositio Mathematica}, 2:463--470, 1935.

\bibitem{DBLP:journals/siamcomp/FederV98}
T.~Feder and M.~Y. Vardi.
\newblock {The Computational Structure of Monotone Monadic SNP and Constraint
  Satisfaction: A Study through Datalog and Group Theory}.
\newblock {\em SIAM J. Comput.}, 28(1):57--104, 1998.

\bibitem{grohe-flum-param}
J.~Flum and M.~Grohe.
\newblock {\em {Parameterized Complexity Theory}}.
\newblock Springer, Berlin, 2006.

\bibitem{DBLP:journals/ipl/FominKN02}
F.~V. Fomin, D.~Kratsch, and J.-C. Novelli.
\newblock {Approximating minimum cocolorings}.
\newblock {\em Inf. Process. Lett.}, 84(5):285--290, 2002.

\bibitem{Greene74}
C.~Greene.
\newblock {An Extension of Schensted's Theorem}.
\newblock {\em Advances in Mathematics}, 14:254--265, 1974.

\bibitem{DBLP:conf/isaac/Guillemot11}
S.~Guillemot.
\newblock {Parameterized Algorithms for Inclusion of Linear Matchings}.
\newblock In {\em ISAAC}, pages 354--363, 2011.

\bibitem{DBLP:conf/isaac/GuillemotV09}
S.~Guillemot and S.~Vialette.
\newblock {Pattern Matching for 321-Avoiding Permutations}.
\newblock In {\em ISAAC}, pages 1064--1073, 2009.

\bibitem{DBLP:conf/swat/HeggernesKLRS10}
P.~Heggernes, D.~Kratsch, D.~Lokshtanov, V.~Raman, and S.~Saurabh.
\newblock {Fixed-Parameter Algorithms for Cochromatic Number and Disjoint
  Rectangle Stabbing}.
\newblock In {\em SWAT}, pages 334--345, 2010.

\bibitem{DBLP:books/aw/Knuth68}
D.~E. Knuth.
\newblock {\em {The Art of Computer Programming, Volume I: Fundamental
  Algorithms}}.
\newblock Addison-Wesley, 1968.

\bibitem{MacMahon}
P.~A. MacMahon.
\newblock {\em {Combinatory Analysis}}.
\newblock London: Cambridge University Press, 1915.

\bibitem{DBLP:journals/jct/MarcusT04}
A.~Marcus and G.~Tardos.
\newblock {Excluded permutation matrices and the Stanley-Wilf conjecture}.
\newblock {\em J. Comb. Theory, Ser. A}, 107(1):153--160, 2004.

\bibitem{MR84}
R.~M\"{o}hring and F.~Radermacher.
\newblock {Substitution decomposition for discrete structures and connections
  with combinatorial optimization}.
\newblock In {\em Algebraic and combinatorial methods in operations research},
  volume~95 of {\em North-Holland Math. Stud.}, pages 257--355, 1984.

\bibitem{DBLP:journals/jcss/Pietrzak03}
K.~Pietrzak.
\newblock {On the parameterized complexity of the fixed alphabet shortest
  common supersequence and longest common subsequence problems}.
\newblock {\em J. Comput. Syst. Sci.}, 67(4):757--771, 2003.

\bibitem{DBLP:conf/stoc/Pratt73}
V.~R. Pratt.
\newblock {Computing Permutations with Double-Ended Queues, Parallel Stacks and
  Parallel Queues}.
\newblock In {\em STOC}, pages 268--277, 1973.

\bibitem{DBLP:journals/jal/RosentiehlT84}
P.~Rosenstiehl and R.~E. Tarjan.
\newblock Gauss codes, planar hamiltonian graphs, and stack-sortable
  permutations.
\newblock {\em J. Algorithms}, 5(3):375--390, 1984.

\bibitem{MR829358}
R.~Simion and F.~W. Schmidt.
\newblock {Restricted permutations}.
\newblock {\em European J. Combin.}, 6(4):383--406, 1985.

\end{thebibliography}
\iffull
\appendix
\newcommand{\Blocks}{\vname{Blocks}}
\newcommand{\point}{\vname{point}}
\newcommand{\cols}{\vname{cols}}
\newcommand{\block}{\vname{block}}
\newcommand{\BigSets}{\vname{BigSets}}

\section{Proof of Theorem~\ref{t:marcus-tardos}}\label{sec:appendix}

We follow the proof technique of \cite{DBLP:journals/jct/MarcusT04}. We show by induction on $p+q$ that: if $M \subseteq [p] \times [q]$ is a point set with no $r \times r$-grid, then $|M| \leq f(r) (p+q-2)$. Clearly, we can assume that $p,q,r \geq 2$. The base case of the induction is when $p+q \leq 2 r^2 (r+1)$. In this case, observe that $\binom{r^2}{r} (p+q-2) \geq (r+1)^2$ as $p,q,r \geq 2$. As $|M| \leq \frac{(p+q)^2}{4}$, we thus have $|M| \leq r^4 (r+1)^2 \leq r^4 \binom{r^2}{r} (p+q-2) = f(r) (p+q-2)$. For the general case, we now suppose that $p+q > 2r^2 (r+1)$.

Let $p' = \lceil \frac{p}{r^2} \rceil$ and $q' = \lceil \frac{q}{r^2} \rceil$. We partition $[p]$ into intervals $I_1,\ldots,I_{p'}$ such that each $I_x$ ($1 \leq x < p')$ has length $r^2$, and we partition $[q]$ into intervals $J_1,\ldots,J_{q'}$ such that each $J_y$ ($1 \leq y < q')$ has length $r^2$. For each $x \in [p'], y \in [q']$, we define the block $B_{x,y} = I_x \times J_y$. From $M$, we define a point set $M' \subseteq [p'] \times [q']$ which contains a point $(x,y)$ iff the block $B_{x,y}$ contains a point of $M$. We say that a block $B_{x,y}$ is \emph{wide} (respectively \emph{tall}) if it contains points of $M$ in at least $r$ different columns (respectively rows).

\begin{lemma} \label{l:no-grid} $M'$ contains no $r \times r$-grid.
\end{lemma}

\begin{proof} Towards a contradiction, suppose that $M'$ contains an $r \times r$-grid, via a gridding $G$ consisting of intervals $I'_1,\ldots,I'_r$ and $J'_1,\ldots,J'_r$. Define the gridding $G'$ of $M$ consisting of intervals $I''_1,\ldots,I''_r$ with $I''_x = \cup_{j \in I'_x} I_j$, and of intervals $J''_1,\ldots,J''_r$ with $J''_y = \cup_{j \in J'_y} J_j$. For every $i,j \in [r]$, we have that $M'$ contains a point $(x',y') \in I'_x \times J'_y$, and thus $B_{x',y'} = I_{x'} \times J_{y'}$ contains a point of $M$. It follows that $M$ contains a point of $I''_x \times J''_y \supseteq B_{x',y'}$, and as this holds for every $x,y \in [r]$ we conclude that $M$ contains an $r \times r$-grid, contradiction.
\end{proof}

\begin{lemma} \label{l:wide-blocks} For every $x \in [p']$, the number of blocks in column $x$ that are wide is less than $r \binom{r^2}{r}$.
\end{lemma}

\begin{proof} Suppose the contrary. For each wide block $B_{x,y}$, suppose that it contains points of $M$ in $r$ different columns $x_1,\ldots,x_r$, and associate to $B_{x,y}$ the set $\{x_1,\ldots,x_r\} \subseteq I_x$. There are at most $\binom{r^2}{r}$ possible such sets, and thus there are $r$ blocks $B_{x,y_1},\ldots,B_{x,y_r}$ ($y_1 < \ldots < y_r$) that are assigned the same subset $S = \{x_1,\ldots,x_r\}$. Set $x_{r+1} = r^2 x + 1$, and define the intervals $I'_1,\ldots,I'_r$ by $I'_i = [x_i,x_{i+1}-1]$ for $i \in [r]$. Next, set $y_{r+1} = q'+1$, and define the intervals $J'_1,\ldots,J'_r$ by $J'_j = \cup_{y_j \leq y < y_{j+1}} J_y$ for $j \in [r]$. These two families of intervals define a $r \times r$-gridding $G$. Observe that for every $i,j \in [r]$, $G(i,j)$ intersects $M$, as $B_{x,y_j}$ contains a point in column $x_i$. We conclude that $M$ contains an $r \times r$-grid, a contradiction.
\end{proof}

\begin{lemma} \label{l:tall-blocks} For every $y \in [q']$, the number of blocks in row $y$ that are tall is less than $r \binom{r^2}{r}$.
\end{lemma}

\begin{proof} Follows by the same proof as Lemma \ref{l:wide-blocks}.
\end{proof}

We are now ready to finish the proof. Let $X_1$ denote the set of wide blocks, let $X_2$ denote the set of tall blocks, and let $X_3$ denote the set of nonempty blocks that are neither wide nor tall. We obtain $|X_1| \leq p' r \binom{r^2}{r}$ by Lemma \ref{l:wide-blocks}, $|X_2| \leq q' r \binom{r^2}{r}$ by Lemma \ref{l:tall-blocks}, and $|X_3| \leq f(r) (p'+q'-2)$ by Lemma \ref{l:no-grid}. As each block contains at most $r^4$ points of $M$, and as each block of $X_3$ contains at most $(r-1)^2$ points of $M$, it follows that:
\begin{align*}
|M| &\leq r^4 |X_1| + r^4 |X_2| + (r-1)^2 |X_3|\\
&\leq r^5 \binom{r^2}{r} (p'+q') + (r-1)^2 f(r) (p'+q'-2)\\
&\leq f(r) (r^2-r+1) (p'+q'-2) + 2 r f(r)
\end{align*}
Now, observe that $p' + q'-2 \leq \frac{p+q}{r^2}$, and thus:
\begin{align*}
|M| &\leq f(r) \frac{r^2-r+1}{r^2} (p+q) + 2 r f(r)\\
&\leq f(r) (p+q) - f(r) \frac{p+q}{r^2}  + 2 r f(r)\\
&\leq f(r) (p+q) - 2 f(r) (r+1) + 2 r f(r) = f(r) (p+q-2)
\end{align*}
Here, we have used that $r \geq 2$ in the second inequality, and we have used that $p+q \geq 2r^2 (r+1)$ in the third inequality. We obtain that $|M| \leq f(r) (p+q-2)$, concluding the proof.\\

\textbf{Implementation.} Following the above proof, we describe a recursive algorithm $\textsc{FindGrid}$\\$(p,q,r,M)$ that takes a point set $M \subseteq [p] \times [q]$ with $|M| > f(r) (p+q-2)$, and finds in $O(|M|)$ time an $r \times r$-grid in $M$. Note that by the assumption on $|M|$ we have $p,q = O(|M|)$. The algorithm assumes that $M$ is described as a list of points, and the resulting grid is described by listing the endpoints of the horizontal and vertical intervals. 

We first describe a subroutine $\textsc{FindBlocks}(p,q,r,M)$ that collects the non-empty blocks of $M'$. The result will be represented by a list $\Blocks$, where each entry $b \in \Blocks$ represents a non-empty block $B_{x,y}$ and holds two fields: $\point(b)$ equal to $(x,y)$; $\cols(b)$ equal to the list of non-empty columns of the block, sorted by increasing order. The subroutine proceeds as follows. First, it arranges the points of $M$ in columns, constructing for each $x \in [p]$ the set $L(x) = \{ z \in M : pr_1(z) = x \}$; this can be performed in $O(|M|)$ time. Second, it scans the columns from left to right, collecting the blocks. For each row $y \in [q']$, we maintain a variable $\block[y]$ pointing to the last created block in row $y$. We initialize all variables $\block[y]$ to $\perp$. When processing column $x \in [p]$, we examine each point $(x,y) \in L(x)$, and in each case: (i) we compute the block $B_{x',y'}$ containing $(x,y)$; (ii) if $\block[y'] = \perp$ or $\block[y']$ is a block $b$ such that $pr_1(\point(b)) < x'$, then we allocate a new block $b$, we set $\block[y']$ to $b$, and we initialize $\point(b)$ to $(x',y')$ and $\cols(b)$ to $\{x\}$; (iii) otherwise, if $b = \block[y']$ then we append $x$ to $\cols(b)$ if it was not already present. The list $\Blocks$ is then returned; it is clear that it holds the desired information, and that its construction takes $O(|M|)$ time.

We now describe a second subroutine $\textsc{FindGridOrReduce}(p,q,r,M)$. The subroutine first calls $\textsc{FindBlocks}(p,q,r,M)$ to obtain the list $\Blocks$. Now, for each column $x \in [p']$, it constructs a list $\BigSets[x]$ as follows. Initially each such list is empty. Then, we examine each block $b$ of $\Blocks$, compute $(x,y) = \point(b)$, test if $|\cols(b)| \geq r$, and if so we obtain $S$ an arbitrary $r$-subset of $\cols(b)$, and add $(y,S)$ to $\BigSets[x]$. For each $x \in [p']$, we determine if there are $r$ entries of $\BigSets[x]$ that have the same second component; if so, we find an $r \times r$-grid $G$ as in the proof of Lemma \ref{l:wide-blocks}, and we return $(yes,G)$. If we find no such grid, then we construct the matrix $M'$ containing the points $\point(b)$ for each block $b \in \Blocks$, and we return $(no,M')$. We claim that this algorithm can be implemented to run in $O(|M|)$ time. First, the construction of the lists $\BigSets$ can be done in time $O(\sum_{b \in \Blocks} |\cols(b)|) = O(|M|)$. Second, for a given $x \in [p']$, consider the time needed to determine if there are $r$ entries of $\BigSets[x]$ that have the same second component. We can do this in time $O(r |\BigSets[x]|)$, by constructing a trie of height $r$ where each leaf is labeled by an $r$-set $S$ together with the set $I$ of indices $y \in [q']$ such that $\BigSets[x]$ contains $(y,S)$; note that the insertion of a new set in the trie takes $O(r)$, and that at the end of the construction we need to look for a leaf whose set of indices $I$ contains at least $r$ elements. Thus, the total time needed for this second step is at most $O(\sum_{b \in \Blocks} |\cols(b)|) = O(|M|)$. Finally, we can construct the reduced matrix $M'$ in $O(|M|)$ time.

To conclude the description of the algorithm, we implement $\textsc{FindGrid}(p,q,r,M)$ as follows. First, we call $\textsc{FindGridOrReduce}$ $(q,p,r,M^t)$, where $M^t = \{ (y,x) : (x,y) \in M \}$. If this call returns $(yes,G')$, we conclude that $G'$ is an $r \times r$-grid in $M^t$, and we return the corresponding $r \times r$-grid in $M$. Otherwise, we call $\textsc{FindGridOrReduce}(p,q,r,M)$. If this call returns $(yes,G'')$, we conclude that $G''$ is an $r \times r$-grid in $M$, and we return it. Otherwise, we obtain $(no,M')$, where $M'$ is the set of points $(x,y) \in [p'] \times [q']$ such that $B_{x,y}$ intersects $M$. As the two calls to \textsc{FindGridOrReduce} answered negatively, we have obtained that $|X_1| \leq p' r \binom{r^2}{r}$ and that $|X_2| \leq q' r \binom{r^2}{r}$, and by the above proof we conclude that $|M'| > f(r) (p'+q'-2)$. We remove some points of $M'$ to obtain $M'' \subseteq M'$ such that $f(r) (p'+q'-2) < |M''| \leq 1.1 f(r) (p'+q'-2)$. This is possible: since $f(r) (p'+q'-2) \geq f(2) \geq 10$, we have $f(r) (p'+q'-2) + 1 \leq 1.1 f(r) (p'+q'-2)$. Finally, we call recursively $\textsc{FindGrid}(p',q',r,M'')$. 

The correctness of the algorithm follows from the above proof, so let us argue about the running time. Consider a call to $\textsc{FindGrid}(p,q,r,M)$ with $|M| > f(r) (p+q-2)$. Assume that the two resulting calls to $\textsc{FindGridOrReduce}$ take time at most $c_1 |M|$, and that the instructions executed inside the call to \textsc{FindGrid} (excluding the function calls) take time at most $c_2 |M|$. Let $c_0 = c_1+c_2$, let $c$ be the solution of $c = c_0+\frac{3.3 c}{r^2}$, and let $c' = \frac{2.2cf(r)}{r^2}$. As $r \geq 2$, it holds that $c$ is positive and that $c \geq c_0$. We show by induction on $p+q$ that the call to $\textsc{FindGrid}(p,q,r,M)$ takes time at most $T(M) \leq c|M|$. If the call issues no recursive call, then it takes time at most $c_0 |M| \leq c |M|$. Suppose now that it issues a recursive call $\textsc{FindGrid}(p',q',r,M'')$ with $f(r) (p'+q'-2) < |M''| \leq 1.1 f(r) (p'+q'-2)$, and that this recursive call takes time at most $c |M''|$ by induction hypothesis. Considering the time taken by the initial call, we obtain:
\begin{align*}
T(M) &\leq c_0 |M| + c |M''|\\
&\leq c_0 |M| + 1.1 c f(r) (p'+q'-2)\\
&\leq c_0 |M| + 1.1 c \frac{f(r)}{r^2} (p+q-2) + c'\\
&\leq c_0 |M| + 3.3 c \frac{|M|}{r^2}\\
&\leq c |M|
\end{align*}
Here we used that $|M''| \leq 1.1 f(r) (p'+q'-2)$ in the second inequality, that $p'+q'-2 \leq \frac{p+q}{r^2}$ in the third inequality, and that $|M| > f(r) (p+q-2)$ and $c' \leq \frac{2.2 c |M|}{r^2}$ in the fourth inequality. As $c$ is bounded by a constant independent of $r$ ($c \leq 5.72 c_0$), we conclude that the running time of $\textsc{FindGrid}(p,q,r,M)$ is $O(|M|)$.

\fi
\end{document}